\documentclass[fleqn,10pt]{./aux/wlscirep}
\usepackage[utf8]{inputenc}
\usepackage[T1]{fontenc}
\usepackage{graphicx}	
\usepackage{color}
\usepackage{amsmath}	
\usepackage{amssymb}	
\usepackage{longtable}  
\usepackage{setspace}
\usepackage{ulem}
\usepackage{newfloat}
\DeclareFloatingEnvironment[name={Extended Data Figure},fileext=lof]{extfigure}
\DeclareFloatingEnvironment[name={Supplementary Figure},fileext=lof]{suppfigure}
\DeclareFloatingEnvironment[name={Supplementary Table},fileext=lof]{suptable}

\newcommand{\target}{J191213.72$-$441045.1} 
\newcommand{\targ}{J1912$-$4410} 
\newcommand{\sun}{_{\odot}} 
\newcommand{\xmmn}{\textit{XMM-Newton}}
\newcommand{\ero}{\textit{eROSITA}}

\newcommand{\warwick}{1}
\newcommand{\saao}{2}
\newcommand{\capet}{3}
\newcommand{\ufs}{4}
\newcommand{\oxford}{5}
\newcommand{\rhodes}{6}
\newcommand{\sarao}{7}
\newcommand{\uj}{8}
\newcommand{\aip}{9}
\newcommand{\up}{10}
\newcommand{\sheffield}{11}
\newcommand{\telaviv}{12}
\newcommand{\ufrgs}{13}
\newcommand{\ing}{14}
\newcommand{\cambridge}{15}
\newcommand{\iac}{16}

\usepackage{lineno}

\title{A 5.3-minute-period pulsing white dwarf in a binary detected from radio to X-rays}

\author[\warwick*]{Ingrid Pelisoli}

\author[\warwick]{T.~R. Marsh}
\author[\saao,\capet,\ufs]{David A.~H. Buckley}
\author[\oxford,\rhodes,\sarao]{I. Heywood}
\author[\saao,\uj]{Stephen. B. Potter}
\author[\aip]{Axel Schwope}

\author[\saao,\capet]{Jaco Brink}
\author[\aip,\up]{Annie Standke}
\author[\capet]{P.~A. Woudt}

\author[\sheffield]{S.~G. Parsons}
\author[\telaviv]{M.~J. Green}
\author[\ufrgs]{S.~O. Kepler}
\author[\warwick,\ing]{James Munday}
\author[\ufrgs]{A.~D. Romero}

\author[\cambridge]{E. Breedt}
\author[\sheffield]{A.~J. Brown}
\author[\sheffield,\iac]{V.~S. Dhillon}
\author[\sheffield]{M.~J. Dyer}
\author[\sheffield]{P. Kerry}
\author[\sheffield]{S.~P. Littlefair}
\author[\sheffield]{D.~I. Sahman}
\author[\sheffield]{J.~F. Wild}

\affil[\warwick]{Department of Physics, University of Warwick, Coventry, CV4 7AL, UK}
\affil[\saao]{South African Astronomical Observatory, PO Box 9, Observatory, 7935, Cape Town, South Africa}
\affil[\capet]{Department of Astronomy, University of Cape Town, Private Bag X3, Rondebosch 7701, South Africa}
\affil[\ufs]{Department of Physics, University of the Free State, PO Box 339, Bloemfontein 9300, South Africa}
\affil[\oxford]{Astrophysics, Department of Physics, University of Oxford, Keble Road, Oxford, OX1 3RH, UK}
\affil[\rhodes]{Department of Physics and Electronics, Rhodes University, PO Box 94, Makhanda 6140, South Africa}
\affil[\sarao]{South African Radio Astronomy Observatory, 2 Fir Street, Observatory 7925, South Africa}
\affil[\uj]{Department of Physics, University of Johannesburg, PO Box 524, Auckland Park 2006, South Africa}
\affil[\aip]{Leibniz-Institut für Astrophysik Potsdam (AIP), An der Sternwarte 16, 14482 Potsdam, Germany}
\affil[\up]{University of Potsdam, Institute for Physics and Astronomy, Karl-Liebknecht-Stra\ss e 24/25, 14476 Potsdam, Germany}
\affil[\sheffield]{Department of Physics and Astronomy, University of Sheffield, Sheffield, S3 7RH, United Kingdom}
\affil[\telaviv]{School of Physics and Astronomy, Tel-Aviv University, Tel-Aviv 6997801, Israel}
\affil[\ufrgs]{Instituto de F\'{\i}sica, Universidade Federal do Rio Grande do Sul, 91501-970 Porto Alegre, RS, Brazil}
\affil[\ing]{Isaac Newton Group of Telescopes, Apartado de Correos 368, E-38700 Santa Cruz de La Palma, Spain}
\affil[\cambridge]{Institute of Astronomy, University of Cambridge, Madingley Road, Cambridge CB3 0HA, UK}
\affil[\iac]{Instituto de Astrof\'{i}sica de Canarias, E-38205 La Laguna, Tenerife, Spain}

\affil[*]{ingrid.pelisoli@warwick.ac.uk}

\begin{abstract}

\end{abstract}

\begin{document}


\flushbottom
\maketitle

\thispagestyle{empty}

\vspace{-1cm}



\doublespacing

\noindent {\bf White dwarf stars are the most common stellar fossils. When in binaries, they make up the dominant form of compact object binary within the Galaxy and can offer insight into different aspects of binary formation and evolution. One of the most remarkable white dwarf binary systems identified to date is AR~Scorpii (henceforth AR~Sco). AR~Sco is composed of an M-dwarf star and a rapidly-spinning white dwarf in a 3.56-hour orbit. It shows pulsed emission with a period of 1.97 minutes over a broad range of wavelengths, which led to it being known as a white dwarf pulsar. Both the pulse mechanism and the evolutionary origin of AR~Sco provide challenges to theoretical models. Here we report the discovery of the first sibling of AR~Sco, \target\ (henceforth \targ), which harbours a white dwarf in a 4.03-hour orbit with an M-dwarf and exhibits pulsed emission with a period of 5.30 minutes. This discovery establishes binary white dwarf pulsars as a class and provides support for proposed formation models for white dwarf pulsars.
}

The white dwarf pulsar AR~Sco is detected over a broad range of wavelengths, from radio \cite{Stanway2018} to X-rays \cite{Takata2018}. The spin-down of its rapidly-rotating white dwarf provides enough energy to power the pulses \cite{Marsh2016}, but the exact driving mechanism is not fully understood. Unlike in neutron star pulsars, where no companion is required, binarity seems to play an important role in AR~Sco's pulses. The observed periodicity of 1.97~min is consistent with a reprocessed frequency, the beat frequency between the 1.95~min spin period of the white dwarf and the 3.56-hour orbital period. This suggests that interaction between the white dwarf and the M-dwarf is behind the pulse mechanism. Proposed models for the origin of emission include the surface or coronal loops of the M-dwarf companion \cite{Katz2017}, the magnetosphere of the white dwarf \cite{Takata2017, PotterBuckley2018}, close to the surface of the white dwarf \cite{Plessis2022}, or through an associated bow shock \cite{Geng2016}.

One of the main challenges to explain AR~Sco is to reconcile the present fast spin-down rate with the rapid spin of the white dwarf. The observed spin period requires previous spin-up via mass accretion. That is because non-interacting main sequence stars slow down their rotation as they age \cite{Barnes2007}, resulting in rotation periods of the order of days for their white dwarf remnants\cite{Hermes2017, Corsico2019}. Only white dwarfs in cataclysmic variables have rotation periods comparable to AR~Sco, which is explained by angular momentum gain via mass accretion from the companion\cite{Patterson1994}. However, whereas the spin-down rate of AR~Sco suggests that a strong magnetic field of 50–100~MG provides the synchronising torque \cite{Katz2017}, the rapid spin can only be achieved with typical mass transfer rates via Roche-Lobe overflow if the magnetic field is much smaller ($\sim 1$~MG)\cite{Lyutikov2020}. With the strong magnetic field, a very large mass transfer rate of up to $\dot{M} \sim 10^{-4}\,\mathrm{M}_\odot \,\mathrm{yr}^{-1}$ would be required to achieve a 1.95~min spin\cite{1979ApJ...232..259G}. This rate is well beyond the ordinary M-dwarf companion seen in AR~Sco, and is $10^5$ times greater than the rates typical of similar binaries\cite{Pala2022}. Accretion via diamagnetic blobs could allow for lower rates\cite{Wynn1995}, but the origin of the strong magnetic field in AR~Sco would remain unexplained.

A solution to this conundrum has recently been put forward \cite{Schreiber2021}. In the proposed model, the white dwarf in AR~Sco was originally non-magnetic, allowing for straightforward accretion-driven spin-up. When crystallisation started to occur in the core of the cooling white dwarf, strong density stratification combined with convection created the conditions for a dynamo, generating the magnetic field \cite{Isern2017,Ginzburg2022}. With a strong enough field, the rapid transfer of spin angular momentum into the orbit causes the binary to detach and mass transfer to briefly cease, leading to a rare system such as AR~Sco. After a few millions of years, the system comes into contact again due to reduced magnetic braking and gravitational radiation, giving origin to a rapidly rotating accreting magnetic white dwarf. As well as addressing the issue of forming a system like AR~Sco, the proposed model also provides a solution to a long-lasting problem in the field of white dwarf binaries: the discrepancy between the fraction of magnetic white dwarfs in detached versus accreting binaries. Strongly magnetic white dwarfs are nearly absent in detached white dwarf binaries \cite{Liebert2015, Parsons2021}, whereas more than one third of those in accreting systems are magnetic \cite{Pala2020}. The dynamo mechanism can naturally explain that, as magnetic accreting white dwarfs are typically old and cool enough to have been crystallised, and they have been spun up by accretion to short periods such that the dynamo effect is intensified. 

An important aspect of this promising rotation- and crystallisation-driven dynamo model is that it suggests that binary white dwarf pulsars like AR~Sco are a possible stage in the evolution of accreting magnetic white dwarfs. Though the timescales in the model cannot be precisely established, given the existence at the time of only one object available to calibrate it, the properties of AR~Sco itself suggest that other binary white dwarf pulsars should exist. First, considering its distance to Earth of only 117~pc, more objects should be expected at larger volumes; second, its spin-down rate suggests that the lifetime in such a stage is of millions of years\cite{Marsh2016}. This model also predicts that the white dwarfs in AR~Sco-like systems should be cool enough to have crystallised, and that their companions should be close to Roche-lobe filling. Finally, based on the observed population of accreting systems, the model also predicts white dwarf pulsars to have orbital periods in the range of three to five hours.

Although a few systems have been proposed as potential white dwarf pulsars\cite{Kato2021a, Kato2021b, Kato2022}, none of them have been confirmed, and as such AR~Sco has remained unique even after six years of its discovery. To address the lack of other systems that could confirm binary white-dwarf pulsars as an evolutionary stage and help to constrain the timescales and predictions of the rotation- and crystallisation-driven dynamo model, we performed a targeted search for binary white-dwarf pulsars. We searched for objects showing observational properties similar to AR~Sco, in particular non-thermal infrared colours, variability, and location in the {\it Gaia} colour-magnitude diagram\cite{Pelisoli23}. A few tens of candidates were identified, over two thirds of which have already been followed-up. Follow-up high-speed photometry with ULTRACAM\cite{ultracam} on the 3.58~m New Technology Telescope (NTT) revealed that one of the candidates, \targ, shows strong pulses with a period of 5.3~min, during which the $g$-band flux increases by up to a factor of four. Independently, \targ\ was detected as an X-ray source with \ero\ during its all-sky surveys \cite{erosita} and identified as a compact binary candidate due to its combined optical (as seen by {\it Gaia}) and X-ray properties. These discoveries prompted further follow-up.

We obtained ULTRACAM photometry on a total of five nights. We also obtained photo-polarimetry on five nights with the High speed Photo-Polarimeter (HIPPO\cite{Potter2010}) mounted on the 1.9~m telescope at the South African Astronomical Observatory (SAAO). Additionally, we retrieved archival photometry from the Transiting Exoplanet Survey Satellite (TESS)\cite{tess}, the Asteroid Terrestrial-impact Last Alert System (ATLAS)\cite{atlas}, the Catalina Real-time Transient Survey (CRTS)\cite{crts}, and the All-Sky Automated Survey for Supernovae (ASAS-SN)\cite{asassn}. Spectroscopy was firstly obtained with the Goodman spectrograph \cite{goodman} at the 4.0~m Southern Astrophysical Research (SOAR) telescope and with the 1.0~m telescope at SAAO, which allowed us to confirm that \targ\ showed similar spectral characteristics to AR~Sco: an optical spectrum displaying a blue continuum added to the red spectrum of an M-type dwarf, with strong Balmer and neutral helium lines in emission (see Extended Data Figure~\ref{fig:spec}). Fast frame-transfer spectroscopy was obtained with the 9.2~m Southern African Large Telescope (SALT), and full orbital coverage was obtained with X-shooter\cite{Vernet2011} at the 8.2~m Very Large Telescope (VLT). Further X-ray data were obtained with \xmmn, after the detection with \ero\ which demonstrated the feasibility of such observations. Radio imaging follow-up observations were carried out using MeerKAT's L-band receivers (856--1712 MHz). Further details on the observations and data reduction are given in the Methods section.

\begin{figure*}[h!]
\centering
\includegraphics[width=0.8\hsize]{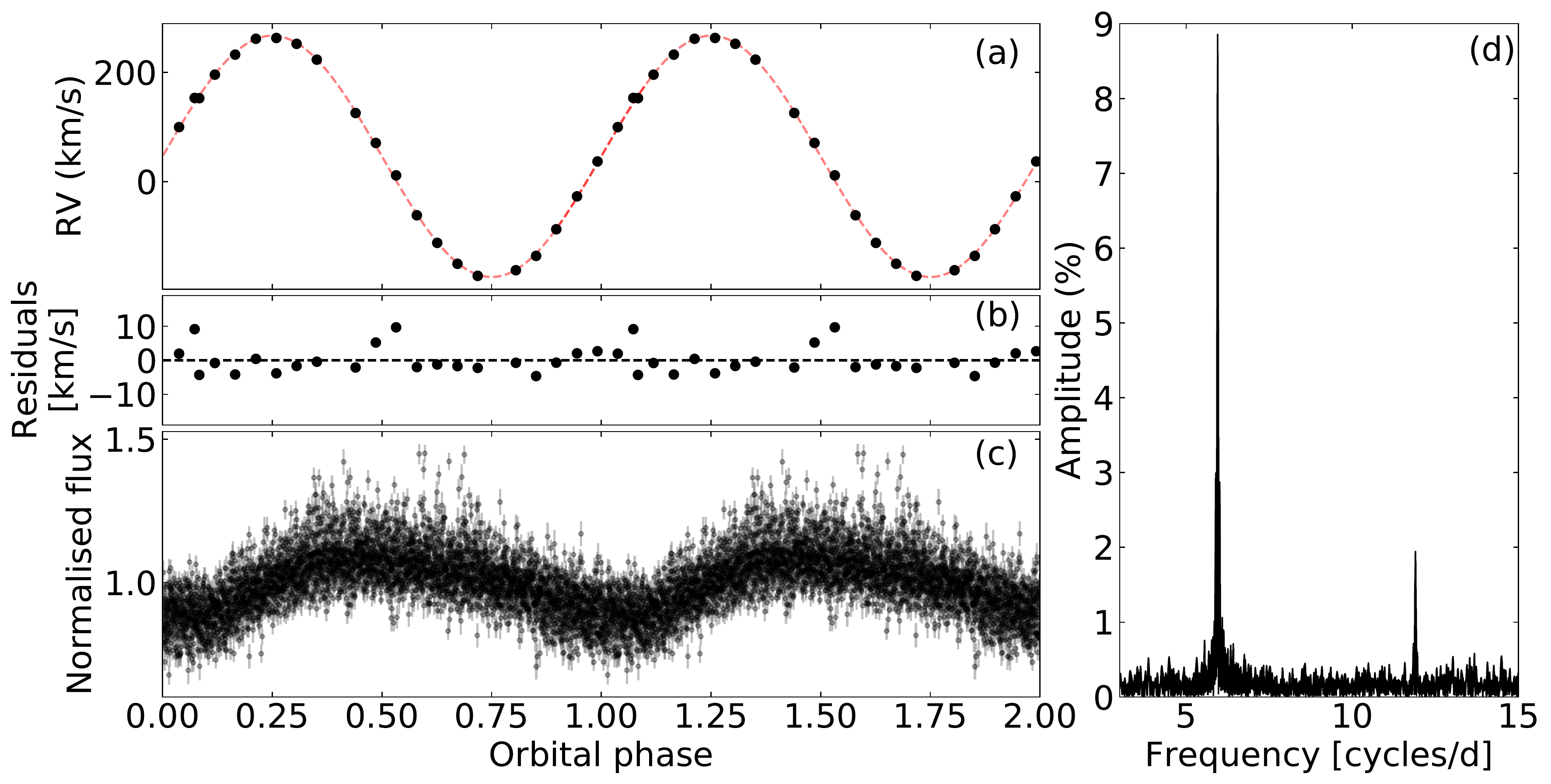}
\caption{{\bf \targ's photometry and radial velocities.} Panel (a) shows the radial velocities obtained from the Na{\sc i} doublet lines (8183 and 8195~\AA) phased to the orbital ephemeris (Eq.~\ref{eq:oeph}). The red dashed line is a sinusoidal fit used to obtain the radial velocity semi-amplitude of the M-dwarf. Uncertainties are of the same order as the symbol size. Panel (b) shows the residuals when the sinusoidal is subtracted (uncertainties are the same as in panel a). Panel (c) shows the TESS data with one-sigma uncertainties folded on the orbital ephemeris. The Fourier transform of the TESS data is shown in panel (d), evidencing the orbital period (strongest peak) and its first harmonic.}
\label{fig:tessRV}
\end{figure*}

The TESS photometry revealed a dominant frequency at $5.948136(13)$~cycles/d. Radial velocities obtained for the M-dwarf from the X-shooter spectra show that this frequency corresponds to the orbital period of the binary, of $0.16811989(36)$~days (Fig.~\ref{fig:tessRV}). The photometric orbital modulation is asymmetric, which cannot be explained by reflection alone. This is likely due to contribution from non-thermal emission, with the asymmetry arising either due to more power being dissipated in the leading face of the M-dwarf, or due to a misaligment between spin and orbit axis, leading to phase-dependent dissipation rate \cite{Katz2017}. The observed radial velocity amplitude depends on the spectral line, reflecting the fact that the emission or absorption features originate in different locations on the M-dwarf. We used the difference between the amplitude measured the from NaI absorption lines (8183 and 8195~\AA), inherent to the M-dwarf photosphere, and the Balmer emission lines, which originate on the irradiated side facing the compact object, to constrain the Roche geometry of the system (details in the Methods section). Additionally, we imposed an upper mass limit equal to the Chandrasekhar mass for the compact object, requiring it to be a white dwarf due to its spin period, which is more than a factor of four longer than any confirmed neutron star pulsar\cite{Caleb2022}. With this upper mass limit, a mass ratio of $q = M_2/M_1 > 0.3$\footnote{We use the subscripts 1 and 2 to refer to parameters describing the white dwarf and the M-dwarf, respectively.} is needed for the origin of both emission and absorption features to fit within the Roche lobe of the M-dwarf (see Extended Data Figure~\ref{fig:roche}). The observed centre-of-mass radial-velocity amplitude of the M-dwarf can only be explained for this range of $q$ if the mass of the white dwarf is $M_1 > 0.32~$M$_{\odot}$. An upper limit to the M-dwarf mass can be obtained by requiring it to fit within its Roche lobe, where it would have a mean density determined by the orbital period\cite{Eggleton1983}. We obtain $M_2 < 0.42$~M$_{\odot}$. The spectral type of the M-dwarf is M4.5$\pm$0.5 (see Methods), suggesting a mass of $\approx 0.3$~M$_{\odot}$, much below the Roche-lobe filling upper limit. This implies that either the M-dwarf is not close to Roche lobe filling, unlike AR~Sco\cite{Pelisoli2022c}, or that the system is seen at low inclination, resulting on a lower observed difference between the velocities of NaI and the Balmer lines due to projection along the line of sight. In fact, we find that our mass constraints imply a maximum inclination for the system of $i = 37^{\circ}$, favouring the latter interpretation (see Extended Data Figure~\ref{fig:roche}, panel b). Combining these constraints with the {\it Gaia} data release 3 distance of $237\pm5$~pc\cite{Bailer-Jones2021}, which allows us to constrain the radius of the M-dwarf from spectral fitting, we estimate the system masses to be $M_1 = 1.2\pm0.2$~M$_{\odot}$ and $M_2 = 0.25\pm0.05$~M$_{\odot}$ (see Extended Data Figure~\ref{fig:masses}), with the companion potentially filling over 90\% of its Roche lobe. The white dwarf temperature cannot be determined due to the lack of contribution of the white dwarf in the optical wavelengths, but the upper limit derived from the lack of visible features is consistent with at least the onset of crystallisation (details in the Methods section).

\begin{figure*}[h!]
\centering
\includegraphics[width=\hsize]{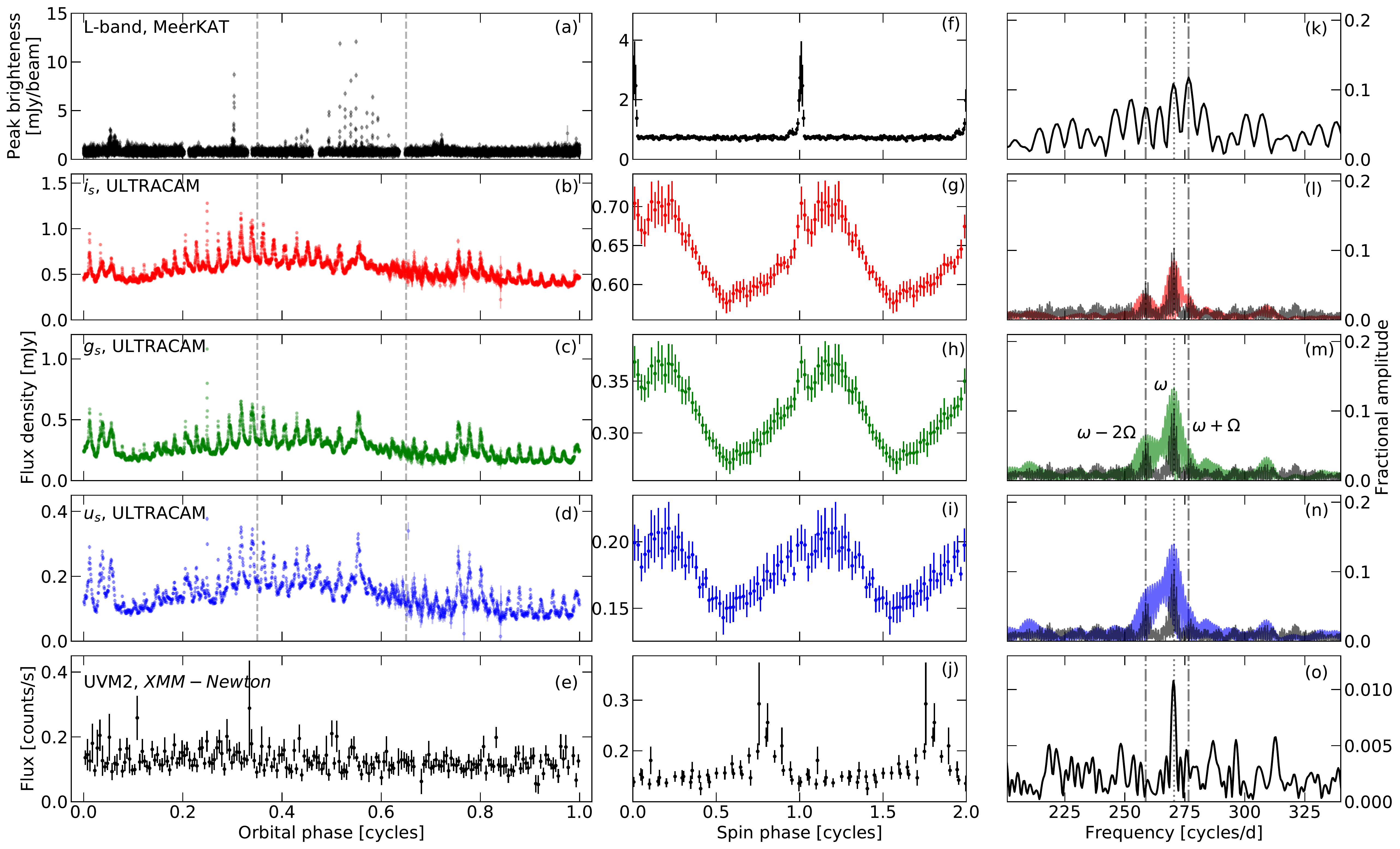}
\caption{{\bf Radio, optical, and X-ray fluxes of \targ.} The panels on the left show high-speed photometry obtained with MeerKAT (a), with ULTRACAM (b--d, filters $i_s$, $g_s$, and $u_s$, respectively), and \xmmn\ (e) as a function of orbital phase. For \xmmn, the full dataset was phase-averaged, whereas for MeerKAT the data covering two orbits were folded with no averaging. For ULTRACAM, we show unfolded data for one orbit. The individual measurement uncertainties are shown in panels (a)-(e), but are sometimes comparable to symbol size.  The dashed vertical lines mark regions of orbital phase that were selected for the spin-phase average pulses shown in the central panels, (f) to (j). For \xmmn, the whole dataset was used. The error bars show the uncertainty on the mean in each phase bin. The panels on the right show the Fourier transform for each dataset, with the main frequency combinations between spin ($\omega$) and orbit ($\Omega$) identified in panel (m). For MeerKAT we used only data around orbital phase 0.5, where the pulses are visible, to calculate the Fourier transform. For ULTRACAM, all data for the five nights are included. We also plot the Fourier transform of the HIPPO data, in which the peaks indicated by dashed lines are better resolved, in the background of panels l--n.
}
\label{phaseFT}
\end{figure*}

\begin{figure*}[h!]
\centering
\includegraphics[width=0.8\hsize]{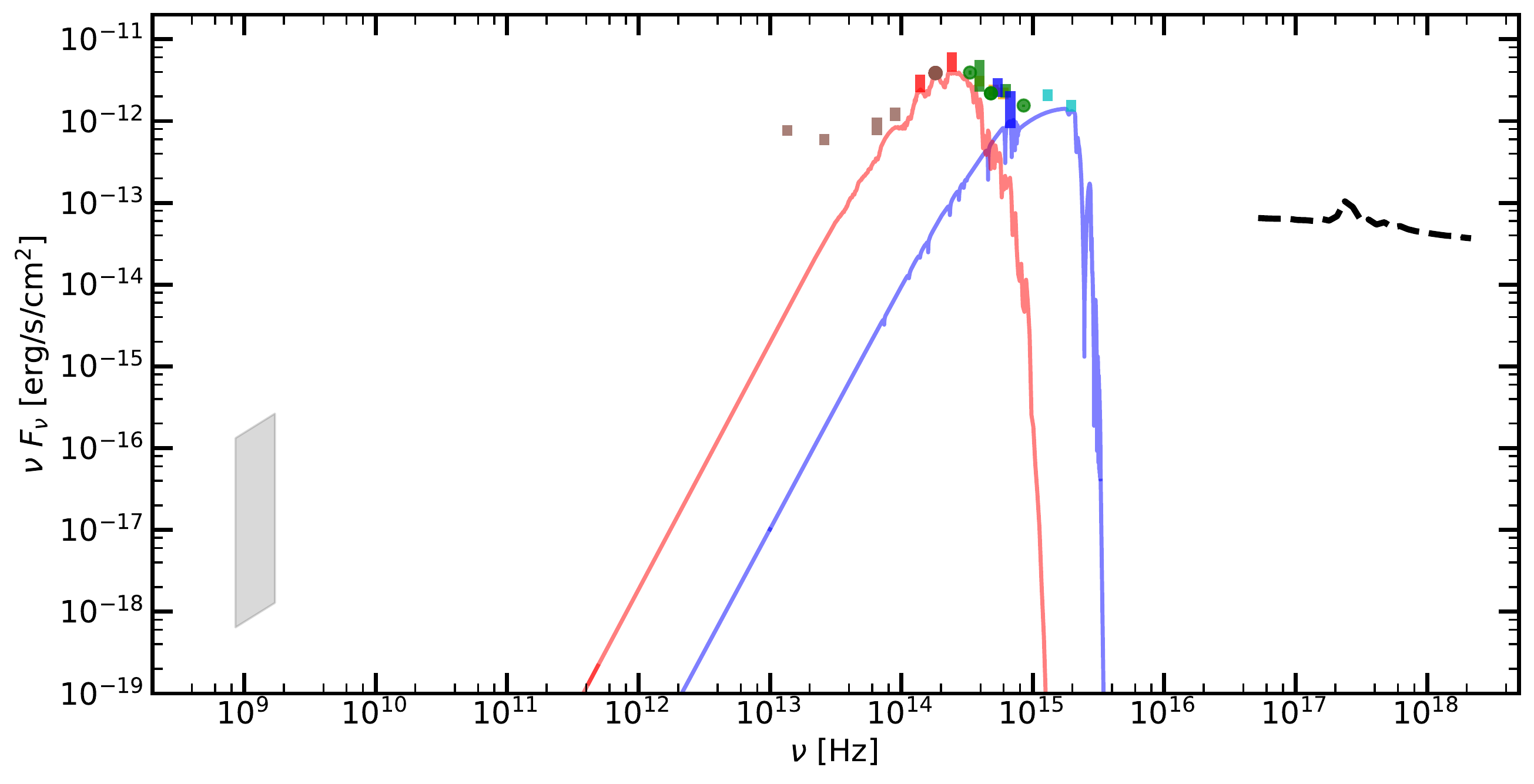}
\caption{{\bf Spectral energy distribution of \targ.} The red and blue lines show model atmospheres assuming parameters close to the constraints that we placed for the M star ($R_2 = 0.3$~R$_{\odot}$, $T_2 = 3100$~K) and white dwarf ($\log~g = 9.0$, and $T_1 = 13000$~K) at the {\it Gaia} distance of $237$~pc. The grey polygon shows the flux and frequency ranges observed in radio. The dashed line models the \xmmn\ spectrum in X-rays, which combines a power law and absorption due to cold interstellar matter. The coloured symbols show the other flux measurements, with vertical bars spanning minimum to maximum values when more than one measurement is available. From left to right, the flux values are W4, W3, W2, W1, H, J, K, z, i, r, V, g, U, u, NUV, and FUV.}
\label{fig:SED}
\end{figure*}

The amplitude spectra computed from our fast-speed photometry reveal several peaks separated by multiples of the orbital frequency (Fig.~\ref{phaseFT}). We interpret the dominant peak of 270.55038(7)~cycles/day as the spin frequency of the white dwarf (more details in the Methods section), corresponding to a spin period of 319.34903(8)~s. Our short baseline of under 5~months reveals no detectable spin period change, which is unsurprising: if \targ\ shows a similar spin-down rate to AR~Sco, the expected change in the spin period over our baseline would be of the order of $10^{-9}$~s, much smaller than the precision in our derived spin period. The pulses are detected in all observed frequencies from radio to X-rays, showing that, like AR~Sco, \targ\ shows a broad spectral energy distribution (SED) that can be orders of magnitudes brighter than the thermal emission from its component stars, particularly at radio and X-ray wavelengths (Fig.~\ref{fig:SED}). Integrating the SED shown in Fig.~\ref{fig:SED} and using the {\it Gaia} distance, we find the bolometric luminosity to be $\sim 10^{33}$~erg/s, well in excess of the total stellar luminosity of $\sim 10^{31}$~erg/s. The excess is even higher than in AR~Sco, potentially suggesting a faster spin-down rate, or another source of energy such as accretion (see below).

\begin{figure*}[h!]
\centering
\includegraphics[width=0.9\hsize]{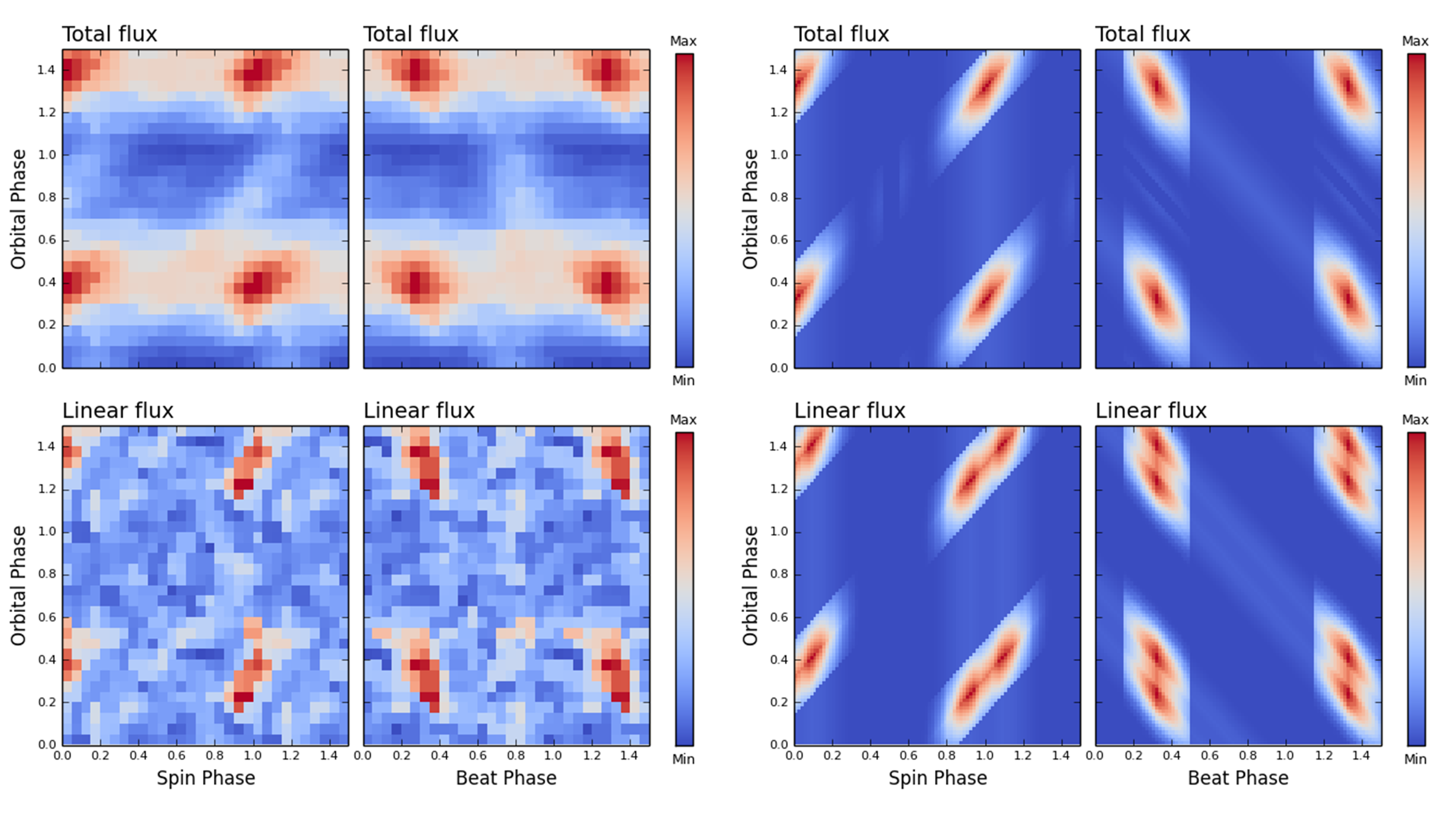}
\caption{{\bf Photopolarimetry of \targ.} The four images on the left show the total intensity (upper two images) and linear polarisation (lower two images). The four images on the right show the model simulations of synchrotron emission from one emission region located close to one magnetic pole of a spinning white dwarf. The images show how the spin and beat (left and right columns, respectively) pulse profiles change with orbital phase. The images represent the averaged phase-folded data of all the photo-polarimetric observations. The total and linear flux have been normalised.}
\label{fig:phot_pol}
\end{figure*}

We interpret the mechanism behind the broad-band pulsed emission to be the same as in AR~Sco: synchrotron radiation. A synchrotron emission source locked to the rotating frame of the white dwarf, which receives an injection of electrons as the magnetic field of the white dwarf sweeps past the companion, can reproduce the observed pulse profile, as illustrated in Fig.~\ref{fig:phot_pol} (details in the Methods section). The electrons are accelerated to relativistic speeds due to magnetic reconnection, resulting in beamed synchrotron emission. The pulsed emission show a complex shape that varies with orbital phase and, to some extent, over time (see Extended Data Figure~\ref{fig:pulses}). At least two components can be identified. First, a narrow component which is particularly dominant in radio, where its full width at half maximum (FWHM) is less than 4~s. This narrow pulse is also seen in the optical, over a much broader second component that dominates both optical and X-rays (see middle panels in Fig.~\ref{phaseFT}). The broader component seen in all wavelengths is likely the result of synchrotron beaming coming at each spin cycle when the magnetic field of the white dwarf sweeps past the companion. The narrow pulses that dominate in radio are reminiscent of neutron star pulsar variability and unlike the broad radio pulses observed for AR~Sco\cite{Stanway2018}. Perhaps the low inclination of \targ\ provides a better view of a magnetic pole, which could be emitting in a similar manner to neutron star pulsars, with very narrow pulses that are directly detected. The phase lag between the two pulses suggests different origins in the system, or at least distinct optical path lengths depending on energy.

\begin{figure*}[h!]
\centering
\includegraphics[width=0.8\hsize]{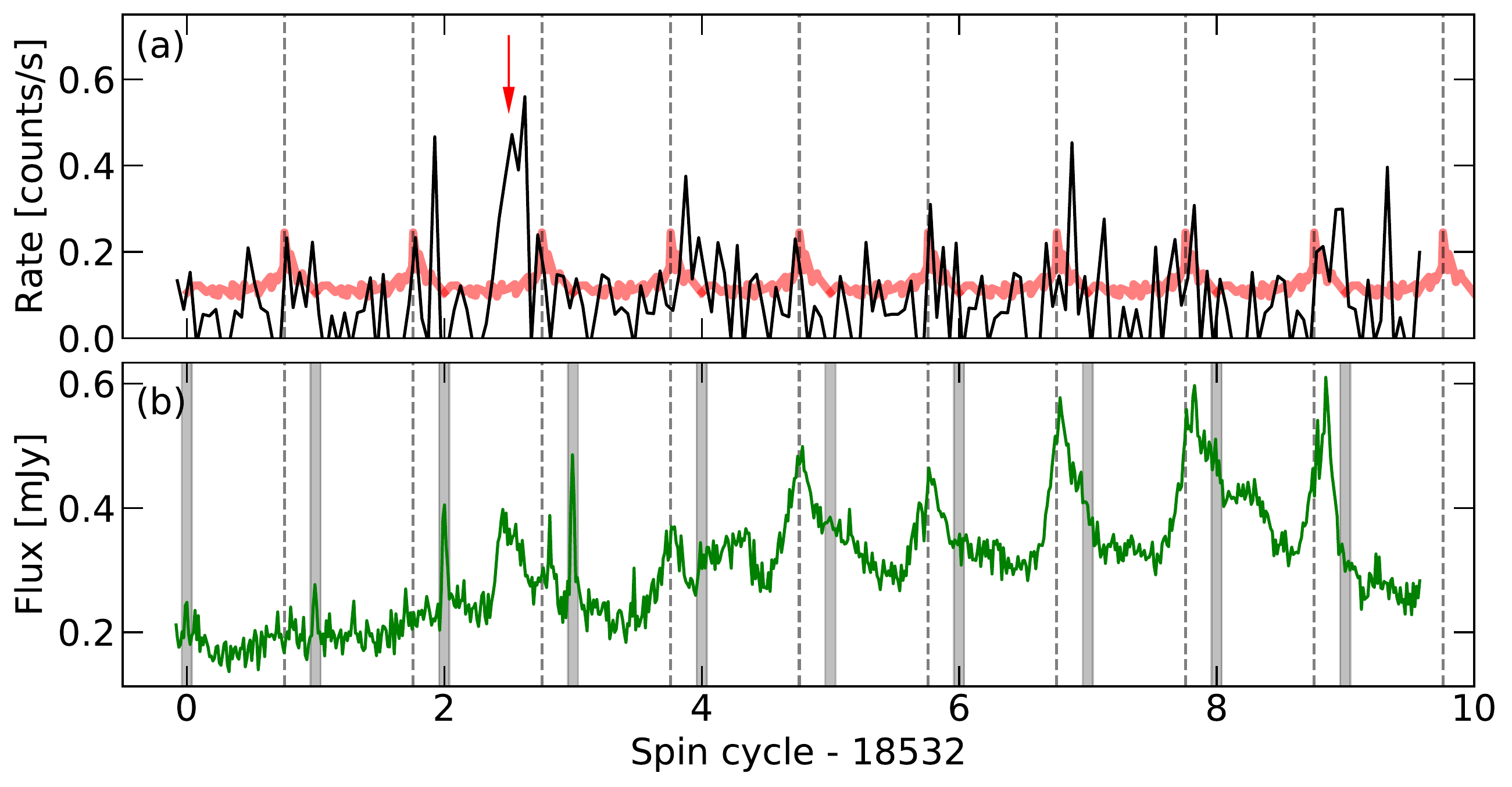}
\caption{{\bf Simultaneous \xmmn\ and ULTRACAM data.} Panel (a) shows in black the \xmmn\ data taken simultaneously with the ULTRACAM $g_s$ data shown in panel (b). The red curve in panel (a) is the spin-phase average \xmmn\ data over the whole observation. The grey shaded areas in panel (b) indicate integer spin cycles, which coincide with the narrow pulses, whereas the dashed grey lines shown in both panels mark the maxima of the broad X-ray pulses. The red arrow indicates a feature that cannot be explained by either the narrow or the wide pulses, which we interpert as a flare.
}
\label{fig:flares}
\end{figure*}

At least one feature seen both in the optical and X-ray data does not correspond to either the narrow or broad pulses (see Figure~\ref{fig:flares}). We interpret this feature as a flare. If the flare were due to chromospheric activity from the M-dwarf, increased amplitude towards the ultraviolet would be expected \cite{Eason1992, Stepanov1995}. That is not what we observe (see Extended Data Figure~\ref{fig:flare_colour}). Instead, we propose that this is an accretion-induced flare, suggestive of mass transfer and subsequent ejection between the M-dwarf and white dwarf (more details in the Methods section). This is in contrast to AR~Sco, which shows no evidence whatsoever of ongoing accretion. Recent models for the evolution of magnetic white dwarfs in close binaries \cite{Schreiber2021} predict accretion to cause the white dwarf spin-up, magnetic field generation, and subsequent detachment of the white dwarf from its companion. The flares in \targ\ suggest that not enough energy has been transferred from spin to orbit to fully detach the system yet, which would suggest it is in a slightly earlier stage of this formation scenario than AR~Sco.

The discovery of \targ\ offers support for the rotation- and crystallisation-driven dynamo model as the origin of magnetic cataclysmic variables. It establishes binary white dwarf pulsars as a class, and provides evidence for residual accretion as predicted by the models\cite{Schreiber2021}. The M-dwarf companion is estimated to be nearly filling its Roche lobe, in agreement with the models and explaining any residual mass transfer. The upper limit on the white dwarf temperature is consistent with the onset of crystallisation, suggesting an emerging magnetic field that provides synchronising torque, which will fully detach the system once enough energy is transferred from the white dwarf spin to the orbit. The orbital period of the system is also in agreement with current model predictions. Assuming AR~Sco and \targ\ are the only two binary white dwarf pulsars in the thin disk suggests a space density of $0.15\pm0.10$~kpc$^{-3}$, taking into account the effective volume given by the thin disk stellar density \cite{juric2008}. This should be regarded as a lower limit, as ongoing searches might reveal other members of this class.


\clearpage

\section*{Methods}

\subsection*{Observations and data reduction}

\noindent {\bf Time-series photometry:} \targ\ was observed with ULTRACAM mounted on the 3.58~m ESO NTT on five nights. ULTRACAM uses dichroic beam splitters that allow simultaneous observations in three filters. We used so-called "super" SDSS filters, whose cut-on/off wavelengths match those of the commonly used Sloan Digital Sky Survey (SDSS) filters, but with a higher throughput \cite{Dhillon2021}. For the first three nights, $u_s, g_s, i_s$ filters were installed, and for the next two $u_s, g_s, r_s$ filters were in place. We used the same exposure time for $g_s, r_s, i_s$ filters, but set it to a factor of 3 or 4 times longer for $u_s$. Frame transfer capabilities were used so that the dead time between exposures was negligible. Supplementary Table~\ref{tab:ucam_log} provides information about each of our runs. We performed bias subtraction and flat field correction, with skyflats taken during twilight, using the HiPERCAM data reduction pipeline\footnote{https://github.com/HiPERCAM/hipercam}. The pipeline was also employed to carry out aperture photometry. We used a variable aperture size set to scale with the seeing estimated from a point-spread function (PSF) fit. The same constant comparison star, Gaia~EDR3~6712712349514963072 ($G = 13.8$), was used in all runs. SkyMapper\cite{skymapper} data were used to calibrate the photometry to an approximate flux scale.

\begin{suptable}[h!]
	\centering
	\caption{{\bf} Log of ULTRACAM observations of \targ.}
	\label{tab:ucam_log}
	\begin{tabular}{ccccc}
		\hline
		\\
		Date       & Start time (UT) & Duration (min) & Filter & Cadence (s) \\
		\\
		\hline
		\\
		           &                 &            &  $u_s$ &        18.6 \\
		2022-04-28 &      $08:49:29$ &         31 &  $g_s$ &         6.2 \\
		           &                 &            &  $i_s$ &         6.2 \\
		\\
		           &                 &            &  $u_s$ &        32.0 \\
		2022-06-07 &      $02:02:02$ &        276 &  $g_s$ &         8.0 \\
		           &                 &            &  $i_s$ &         8.0 \\
		\\
		           &                 &            &  $u_s$ &         9.1 \\
		2022-06-08 &      $05:21:30$ &        243 &  $g_s$ &         3.0 \\
		           &                 &            &  $i_s$ &         3.0 \\
    	\\
		           &                 &            &  $u_s$ &        12.1 \\
		2022-09-18$^{\star}$ &      $03:20:35$ &        51  &  $g_s$ &         4.0 \\
		           &                 &            &  $r_s$ &         4.0 \\
		\\
		           &                 &            &  $u_s$ &         8.5 \\
		2022-09-24 &      $00:30:08$ &       237  &  $g_s$ &         2.8 \\
		           &                 &            &  $r_s$ &         2.8 \\
		\\
		\hline
            $^{\star}$ Simultaneous with \xmmn.
	\end{tabular}
\end{suptable}

\noindent {\bf Photopolarimetry:} Photopolarimetry was performed with HIPPO\cite{Potter2010} on the 1.9-m telescope of the South African Astronomical Observatory during the nights of 2022 June 29 and 30, 2022 July 3--5, as detailed in Supplementary Table~\ref{tab:hippo_log}. Observations were through a clear filter ($3500-9000$~\AA) defined by the response of
the two RCA31034A GaAs photomultiplier tubes. Observations of polarised and non-polarised standard stars were made during the course of the observation run in order to calculate the waveplate position angle offsets, instrumental polarisation, and efficiency factors. The photometry is not
absolutely calibrated and is instead given as total counts minus the background-sky counts. \targ\ was observed for a total of $\sim 32$\ hours. 

\begin{suptable}[h!]
	\centering
	\caption{{\bf} Log of HIPPO observations of \targ. The cadence was 5.0~s for all runs.}
	\label{tab:hippo_log}
	\begin{tabular}{cccc}
		\hline
		\\
		Date       & Start time (UT) & Duration (min)\\
		\\
		\hline
		\\
        2022-06-29 & 20:09:09   & 477.3 \\
        \\
        2022-06-30 & 20:27:09   & 460.8 \\
        \\
        2022-07-03 & 23:01:09   & 274.5 \\
        \\
        2022-07-04 & 19:53:34   & 252.1 \\
        \\
        2022-07-05 & 20:14:24   & 462.0 \\
        \\
        \hline
	\end{tabular}
\end{suptable}

\noindent {\bf Spectroscopy:} Two exploratory spectra to confirm \targ's similarity with AR~Sco were obtained with SOAR on 2022 August 28 during time obtained for project SO2022A-016. The data were reduced using {\tt iraf}’s {\tt noao} package. Upon confirmation of that \targ's had spectral features like those seen in AR~Sco, we obtained Director's Discretionary Time (DDT) with X-shooter (proposal 109.24EM) to cover one full orbital period. The observations were carried out on 2022 July 24, between 03:30:06 and 08:00:20 UT. We used a 1~arcsec slit for the UVB arm (300-559.5~nm, $R = 5400$), and 0.9~arcsec slits for the VIS (559.5-1024~nm, $R = 8900$) and NIR (1024-2480~nm, $R= 5600$) arms. The exposure time in the UVB arm was set to a fifth of the spin period (63.6~s) to sample the spin variability, and we obtained 168 exposures. In the VIS arm, the exposure was set to twice the spin period (636.4~s) to average out the effect of the spin variability, as our main interest with the VIS was to characterise the M-dwarf companion. 21 exposures were obtained. Finally, in the NIR arm the exposure was equal to the spin period (318.2~s) and we obtained 42 exposures. 2x2 binning was used to reduce the readout time, which was of 28~s in UVB, 34~s in VIS and 8.2~s in NIR. Automatic flexure compensation (AFC) exposures were obtained every 1.5~h. The X-shooter data were reduced using the {\tt xsh\_scired\_slit\_stare} routine in the ESO Recipe Execution Tool (EsoRex), and telluric line removal was performed with {\tt molecfit} \cite{molecfit1,molecfit2}.

We also obtained medium-resolution (5.7\AA) time resolved spectra over the wavelength range 4060--7120~\AA\ of \targ\ on 2022 June 26 using the Robert Stobie Spectrograph (RSS) \cite{Burgh2003,Kobulnicky2003,Smith2006} on the Southern African Large Telescope (SALT) \cite{Buckley2006}. Two observations of 3000 s were obtained during both the rising (east) and setting (west) tracks, commencing respectively at 20:10:54 UTC and 02:37:53 UTC.  Frame transfer mode was used, with 60 repeat 50 s exposures for each observation, with no dead-time. 

The spectra were reduced using the \texttt{PySALT} package \cite{Crawford2010}, which includes bias subtraction, flat-fielding, amplifier mosaicing, and a process to remove cosmetic defects. The spectra were wavelength calibrated, background subtracted, and extracted using standard IRAF\footnote{IRAF is distributed by the National Optical Astronomy Observatory, which is operated by the Association of Universities for Research in Astronomy (AURA) under cooperative agreement with the National Science Foundation (NSF).} procedures. We obtained a relative flux calibration of all spectra using the spectrophotometric standard star Feige 110. The frame-transfer observations were used to create trailed spectra in order to investigate emission line variability. 

\noindent {\bf X-rays:} The original X-ray detection of \targ\ was made during eRASS1, the first X-ray all sky survey with \ero\ on the Spektrum-Roentgen-Gamma mission \cite{erosita,Sunyaev2021}. The eRASS1 proto-catalogue internal to the collaboration\footnote{The full catalogue will be published by Merloni et al., in preparation.}, which was produced with processing version c947\cite{brunner+22}, lists the detection of \targ\ at RA$=288.05807$, DEC$=-44.17913$. Source detection was performed on an image using photons in the $0.2 - 2.3$\,keV band. The source was found at a rate of $0.35 \pm 0.07$~s$^{-1}$, which corresponds to a flux of $(3.3 \pm 0.7) \times 10^{-13}$\,erg cm$^{-2}$ s$^{-1}$. 

Detailed follow-up was obtained with \xmmn\ following a DDT request. \xmmn\ observed the field of \targ\ on 2022 Septemper 17/18 for a total of 43~ks. For approximately 50~min, ULTRACAM observations were carried out simultaneously with \xmmn. The EPIC X-ray cameras were operated in full frame mode, and the optical monitor was used in fast imaging mode (time resolution 0.5~s) with the UVM2 filter (effective wavelength 231~nm). The original data were reduced with the latest version of the \xmmn\ SAS (SAS 20.0) using the most recent calibration files. Light curves and spectra were produced using the signal in concentric annuli around the source position to correct for background contamination. The mean rate in EPIC-pn was 0.0825~s$^{-1}$ ($0.2 - 10$\,keV). 

The mean X-ray spectrum was analysed with XSPEC (version 12.12.0) \cite{arnaud+96}. An emission model as the sum of a power law (power law index $2.14 \pm 0.11$) and a thermal component ($kT = 1.24 \pm 0.11$\,keV) modified with absorption due to cold interstellar matter (N$_{\rm H} = (5\pm 2) \times 10^{20}$\,cm$^{-2}$) reflected the data satisfactorily. The best-fit model parameters yielded the model shown in Fig.~\ref{fig:SED}.  A full description of the X-ray data analysis including data from all \ero\ surveys and the \xmmn\ data is presented separately \cite{Schwope23}.

\noindent{\bf Radio:} An exploratory observation of J1912$-$4410 was made with MeerKAT's L-band receivers (856--1712 MHz) for one hour on 22 June 2022 (block ID 1655920939). Following the detection of pulses, a longer observation was made on 26 June 2022 (block ID 1656263834). The total on-target time was 7.55 hours, with the correlator configured to deliver 2~s integrations, which defines the minimum imaging timescale. Instrumental bandpass and delay corrections were derived from scans of the primary calibrator PKS B1934$-$638, and time dependent complex gains were derived from scans of the secondary calibrator J1830$-$3602. The {\tt casa}  package \cite{casa2022} was used for reference calibration and flagging of the calibrator scans. These gain corrections were then applied to the target field, which was flagged using the {\tt tricolour} software \cite{hugo2022}. The target data were imaged with {\tt wsclean} \cite{offringa2014}, and self-calibrated using {\tt cubical} {\tt kenyon2018}. The data reduction scripts\cite{heywood2020} that perform the processing up to this point are available online\footnote{{\tt oxkat}, v0.3, https://github.com/IanHeywood/oxkat}, and contain detailed lists of all the parameters used.

Following self-calibration and a final round of deconvolution, the resulting set of spectral clean components (excluding the position of J1912$-$4410 itself) were inverted into a set of model sky visibilities and subtracted from the data. This residual visibility set was then imaged using {\tt wsclean}, producing an image for every 2~s timeslot in the observation (13,552 snapshot images in total). We experimented with deriving an in-band spectral index using the strongest pulses, but found considerable scatter between pulses, with values ranging from -4.4 to -0.8 for different pulses and uncertainties reaching 40\%. The median suggests a steep negative spectral index of $\approx -3$, but higher signal-to-noise ratio than currently available is required to obtain robust and conclusive results.

\noindent {\bf Archival data:} \targ\ was observed by TESS during sectors 13 and 27, with a cadence of 30~min and 10~min, respectively. We downloaded postcards and performed aperture photometry using {\tt eleanor} in a custom script\footnote{https://github.com/ipelisoli/eleanor-LS}. We also downloaded photometry taken with the cyan ($c$) and orange ($o$) filters from the ATLAS archive\footnote{https://fallingstar-data.com/forcedphot/}. Data were also available in CRTS and ASAS-SN, but due to the target's relative faintness we could not identify any of the periodic variability seen in other data in CRTS or ASAS-SN data, which we therefore employed no further.

\subsection*{The companion subtype and the orbital ephemeris}

To determine the spectral subtype of the companion, we used the VIS arm of the X-shooter spectra. To minimise contribution from the white dwarf, both direct and due to its irradiation on the M-star, we started by deriving radial velocities for the H$\alpha$ emission line by fitting it with a Gaussian profile. H$\alpha$ should trace the irradiated face of the companion (as later confirmed, see the section ``Doppler tomography and the origin of emission"), hence by fitting the velocities with a sinusoidal we were able to estimate the point at which the companion is at its closest approach to Earth, when contribution from the white dwarf is minimised. The spectrum closest to inferior conjunction was then employed for spectral subtype determination. We fitted the observed spectrum using M-dwarf templates obtained with the same X-shooter configuration \cite{Parsons2018} as our observations. The template fluxes were scaled by a factor $\alpha$ and combined with a smooth continuum to account from any extra flux in addition to the M-star. This was parameterised by $\exp(a_1 + a_2\lambda)$ to ensure positivity\cite{Marsh2016}. To focus on the M-dwarf contribution as well as avoid noisy regions, we only fit the spectrum between 6800 and 9250~\AA. Additionally, we masked the CaII emission triplet lines around 8500~\AA. Similar values of $\chi^2$ were obtained for the M4, M4.5, and M5 templates, hence we concluded the companion to be of type M$4.5\pm0.5$.

Once the spectral type was determined, we proceeded to estimate the radial velocity of the M-dwarf by cross-correlating the NaI 8200~\AA\ doublet absorption lines, which should more closely trace the centre of mass than the emission lines, with a spectral template. We found the M4.0 template to provide a better fit in this region, and therefore used it in the cross-correlation. We normalised the region within 8140--8240~\AA\ by the continuum using a first order polynomial, and then subtracted a first-order spline fit to the normalised continuum. The same procedure was applied to the 21 VIS spectra and the template, which were then binned to the same velocity scale and cross-correlated. The obtained radial velocities are listed in Supplementary Table~\ref{tab:rvs}.

\begin{suptable}[h!]
	\centering
	\caption{{\bf Radial velocities obtained for \targ.} Radial velocities and one-sigma uncertainties obtained for the M-dwarf in \targ\ using the NaI 8200~\AA\ doublet absorption lines.}
	\label{tab:rvs}
	\begin{tabular}{cccc}
		\hline
		BJD(TDB)  & RV (km/s) & $\sigma_{\mathrm{RV}}$ (km/s) \\
		\hline
2459784.65911 & 153.2 & 3.6 \\ 
2459784.66687 & 195.8 & 3.5 \\ 
2459784.67463 & 232.4 & 3.6 \\ 
2459784.68259 & 261.5 & 4.0 \\ 
2459784.69042 & 262.9 & 4.1 \\ 
2459784.69818 & 252.2 & 4.5 \\ 
2459784.70593 & 223.3 & 4.4 \\ 
2459784.72079 & 125.6 & 5.0 \\ 
2459784.72855 & 70.8 & 4.6 \\ 
2459784.73632 & 11.4 & 4.6 \\ 
2459784.74427 & -61.7 & 4.6 \\ 
2459784.75208 & -112.2 & 4.5 \\ 
2459784.75984 & -150.5 & 3.9 \\ 
2459784.76760 & -172.5 & 4.0 \\ 
2459784.78221 & -162.3 & 3.9 \\ 
2459784.78997 & -136.0 & 4.7 \\ 
2459784.79771 & -87.2 & 5.0 \\ 
2459784.80566 & -27.0 & 5.2 \\ 
2459784.81350 & 36.8 & 4.8 \\ 
2459784.82127 & 99.7 & 4.4 \\ 
2459784.82901 & 152.7 & 4.8 \\ 
        \hline
	\end{tabular}
\end{suptable}

To determine the orbital ephemeris, the obtained radial velocities were combined with an orbital period measurement from TESS data, whose time span and continuous coverage is ideal for precisely determining the orbital frequency. The Fourier transform of the TESS data showed a strong peak near 5.95~cycles/d, with the first harmonic also clearly visible (see Fig.~\ref{fig:tessRV}). We interpreted this as the orbital frequency, which is in agreement with the observed radial velocity variability and within the orbital period range predicted for white dwarf pulsars\cite{Schreiber2021}.

Reflection alone cannot explain the observed photometric orbital modulation, which is asymmetric and shows amplitude higher than the sub-percent that would be expected for a reflection effect for the system's estimated parameters at low inclination (which we find to be the case for \targ, as discussed in the next section). The larger amplitude can be explained by contribution from non-thermal emission when the rotational energy of the white dwarf is dissipated by interaction of its magnetic field with the M dwarf’s. The asymmetry, also observed in AR~Sco, has two proposed explanations \cite{Katz2017}. The first is that the power dissipated by interaction between the star’s magnetic fields is greater on the leading face of the M-dwarf, where shock occurs, than on its trailing face. The other possibility is that the spin axis of the white dwarf is misaligned with the orbit, which would cause the dissipation rate in the M dwarf to vary with orbital phase. A consequence of the latter is that precession of the spin axis would make the orbital phase of the maximum to drift, which has not been observed for AR~Sco\cite{Littlefield2017}, but cannot be ruled out given that the precession period can be of up to several hundreds of years.

The system's orbital period and its uncertainty were determined via bootstrapping, using a Fourier model with two sine terms (one on the fundamental frequency, and one on the first harmonic) to fit the TESS data. We obtained $P_\mathrm{orb} = 0.16811989(36)$~days, corresponding to $\Omega = 5.948136(13)$~cycles/d. To determine the reference epoch (phase 0) of the ephemeris, defined here as the inferior conjunction of the M-dwarf, we fitted the radial velocities with
\begin{eqnarray}
V_R &=& \gamma + K_2 \sin[2\pi(t-T_0)\Omega],
\end{eqnarray}
where $\gamma$ is the systemic velocity, $K_2$ the radial velocity semi-amplitude of the M-dwarf, $t$ the mid-exposure times of each spectrum yielding a measurement, $T_0$ the reference time of inferior conjunction, and $\Omega$ is the orbital frequency, which was fixed at the value obtained from the TESS data. Using bootstrapping to determine uncertainties, we obtained $\gamma = 46.2\pm1.0$~km/s, $K_2 = 220.9\pm1.1$~km/s, and $T_0 = 2459784.98308(19)$, thus the orbital ephemeris of \targ\ is:
\begin{eqnarray}
BJD(TDB) &=& 2459784.98308(19) + 0.16811989(36) E,
\label{eq:oeph}
\end{eqnarray}
where $E$ is an integral cycle number, and BJD is the barycentric Julian date (in TDB scale).

\subsection*{Constraining the stellar masses}

From the value of $K_2$ and the orbital period, we can use Kepler's third law to calculate the mass function:
\begin{eqnarray}
f_M &=& \frac{M_1^3 \sin ^3 i}{(M_1 + M_2)^2} \; = \; \frac{P_{\mathrm{orb}} K_2^3}{2\pi G} \; = \; (0.1879 \pm 0.0027)~M_{\sun}.
\label{eq:massf}
\end{eqnarray}
This equation, in which $i$ is the orbital inclination, sets a lower limit to the mass of the unseen compact object, met for $M_2 = 0$ and $i = 90^{\circ}$. This only stands if our assumption that the obtained radial velocities indeed trace the centre of mass of the M-dwarf is correct. As we detected no systematic deviation from a sinusoidal in the residuals (see Fig.~\ref{fig:tessRV}), as would be expected if irradiation caused weakening of the absorption lines on the side of the cool star facing the compact companion, our assumption seems to be correct.

An independent constraint on the system's masses can be obtained from the difference between the semi-amplitudes derived from the emission lines (tracing the irradiated face) compared to the absorption lines (tracing the centre of mass), shown in Supplementary Table~\ref{tab:rv_amps}. The simple assumption that all measurements have to be within the M-dwarf's Roche-lobe results in $q > 0.1$, as illustrated in panel (a) of Extended Data Fig.~\ref{fig:roche}. Requiring that star 1 is a white dwarf can provide a tighter constraint. Given the spin period of 5.3~min, too long for a neutron star but completely consistent with white dwarfs in magnetic cataclysmic variables\cite{Gaensicke2005}, we consider this to be a fair assumption. As shown in panel (b) of Extended Data Fig.~\ref{fig:roche}, this results in $q > 0.3$. This can be combined with Equation~\ref{eq:massf} to obtain a tighter constraint on $M_1$. Equation~\ref{eq:massf} implies
\begin{eqnarray}
M_1 = f_M (1+q)^2/\sin ^3 i.
\label{eq:m1_q}
\end{eqnarray}
Given the lower limit for $q$ and upper limit for $\sin ^3 i$ at $i = 90^{\circ}$, we obtain $M_1 > 0.32$~M$_{\sun}$ and $M_2 > 0.095$~M$_{\sun}$.

\begin{suptable}[h!]
	\centering
	\caption{{\bf Radial velocity semi-amplitudes for different lines}}
	\label{tab:rv_amps}
	\begin{tabular}{cc} 
		\hline
		Line       &  Semi-amplitude (km/s) \\
		\hline
            H$\alpha$ & $177\pm5$  \\
            H$\beta$ & $155.5\pm1.1$  \\
            H$\gamma$ & $157.7\pm1.0$ \\
            NaI doublet & $220.9\pm1.1$ \\
            \hline
	\end{tabular}
\end{suptable}

We can also obtain an upper limit on the M-dwarf mass. For a Roche-lobe filling star, there is a tight relationship between orbital period and mean density $\rho$ \cite{Eggleton1983}:
\begin{eqnarray}
\rho [\mathrm{g/cm}^3] &=& (0.43/P_\mathrm{orb}[\mathrm{days}])^2.
\end{eqnarray}
Because the mean density of a main-sequence star decreases with increasing mass, this corresponds to an upper limit. Assuming a semi-empirical mass-radius relationship\cite{Brown2022}, which takes the M-dwarf inflation problem\cite{Kesseli2018} into account, results in $M_2 < 0.42$~M$_{\sun}$. Combining the limit given by the M-dwarf filling its Roche lobe with the observed $K_2$ values also implies a maximum inclination at which star 1 is a white dwarf. Higher inclinations would require a more compact M-dwarf to explain the observed $K_2$ difference, which in turn requires the compact object to have a higher-mass so that the M-dwarf still fills the Roche lobe. The maximum system inclination is $i < 37^{\circ}$, as demonstrated in Extended Data Fig.~\ref{fig:roche}.

Equation~\ref{eq:m1_q} can also be interpreted as a relationship between $M_1$, $M_2$, and inclination. With the limits derived above, useful constraints can be obtained. This is shown in Extended Data Figure~\ref{fig:masses}. The upper limit on inclination requires the white dwarf mass to be above $\approx 1.0$~M$_{\sun}$. To further constrain the masses, we can rely on an estimate for $M_2$ obtained from an estimate of the stellar radius. The distance to \targ\ is well constrained by the {\it Gaia} parallax to $d = 237\pm5$~pc. Hence, when fitting the observed spectrum to models, the previously mentioned scaling factor $\alpha$, which corresponds to $(R_2/d)^2$, provides a radius estimate. Combining this with a mass-radius relationship gives $M_2$. We fitted NextGen models \cite{Hauschildt1999} to the M-dwarf spectra via $\chi^2$ minimisation using the same wavelength region as employed to obtain the M-dwarf subtype. We kept the $\log~g$ fixed at 5.5 rather than free, given that the effects of gravity and rotation on the line widths cannot easily be disentangled. We obtained $T_2 = 3\,100\pm100$~K and $R_2 = 0.23\pm0.02$~R$_{\odot}$. Using the same mass-radius relationship as above results in $M_2 = 0.26\pm0.02$~M$_{\odot}$. The quoted uncertainties are statistical only. For $R_2$ and $M_2$, the M-dwarf inflation problem suggests that the uncertainty is of the order of 15\%. The irradiation of the M-dwarf by the white dwarf, which can also cause inflation, leads to similar uncertainty\cite{Knigge2011}. The derived $M_2$ value implies a white dwarf mass of $\approx 1.2$~M$_{\sun}$. To take into account the large systematic uncertainties in $M_2$, we adopt the mass values of $M_1 = 1.2\pm0.2$~M$_{\sun}$ and $M_2 = 0.25\pm0.05$~M$_{\sun}$ quoted in the main text. These masses imply $q \sim 0.21$, and hence a Roche-lobe radius of $\sim 0.25$~R$_{\odot}$\cite{Eggleton1983} for the M-dwarf, close to our estimated radius, suggesting that the M-dwarf is near Roche-lobe filling, as inferred for AR~Sco\cite{Pelisoli2022c} and as predicted for a white dwarf pulsar configuration\cite{Schreiber2021}.

\subsection*{The spin and beat frequencies}

The ground-based fast photometry obtained with ULTRACAM and HIPPO was used to determine the spin frequency of \targ. This identification is not as straightforward as the orbital period. Fourier transform of the ULTRACAM and HIPPO data showed three main peaks separated by multiples of the orbital frequency. We also analysed light curves derived from the continua and emission lines of observed spectra (see Supplementary Figure~\ref{fig:linesFT}), which show in turn only one resolved peak. The lines originate in the irradiated face of the companion, hence their variability would likely reflect the reprocessed or beat frequency. The resolution of the spectral Fourier transforms is, however, not high enough to separate spin and beat frequencies. We ultimately relied on the modelling of the photopolarimetry (see Section "Photopolarimetry and modelling of the emission") to determine whether the dominant frequency is the spin or beat, concluding that the data can be better explained by the models if the former is the dominant frequency. Therefore we interpret the two main peaks as $\omega$ and $\omega - 2\Omega$, with $\omega + \Omega$ also being marginally detected. The absence of the beat frequency is somewhat puzzling, although 2($\omega - \Omega$) is detected (Supplementary Figure~\ref{fig:FTbeat}). This could be a consequence of low inclination, such that reprocessed radiation from only one pole is detected. 

To determine the spin ephemeris, we measured pulse arrival times from both the HIPPO and the ULTRACAM data. We first estimated $\omega$ and the time of maxima $T_0^{\omega}$ values by fitting a cosine function to the ULTRACAM $g_s$ data, after subtracting the orbital modulation. This trial ephemeris was used to estimate the approximate pulse arrival times and cycle numbers expected from each dataset. To refine this estimate, we cross-correlated the data around each expected peak with a Gaussian function, and found the time of maxima by locating the maxima of the cross-correlation function using the Newton-Raphson method.

This procedure of estimating times of maxima given trial ephemeris and subsequently fitting the obtained values was repeated until the trial and fitted ephemeris showed no significant change. The $\sigma$ width of the Gaussian was fixed at a value of 15~s, which we found to yield a good balance between identified pulses and fit residuals (i.e. a large number of pulses was obtained, and the identifications were good enough that the residuals to the fit were not increased. Increasing the value of $\sigma$ resulted in lower residuals, but fewer identified pulses, whereas decreasing the value increased the number of identified pulses, but also increased residuals).

Next we fitted the obtained cycle numbers and time of maxima assuming linear ephemeris. The residuals of the fit are modulated with orbital phase, as illustrated in Supplementary Figure~\ref{fig:omc}. The semi-amplitude of the modulation is of the order of 15~s, significantly in excess of the $\lesssim 1.5$~s that could be explained by the difference in light travel time throughout the orbit. This behaviour likely arises due to the varying contribution of beat, spin and their harmonics to the pulse shape throughout the orbit, as seen in AR~Sco \cite{Stiller2018,Gaibor2020,Pelisoli2022c}, and suggests that the line-of-sight geometry plays a role in the detected emission. To minimise the effect of this in the ephemeris determination, we fitted the orbital modulation with a Fourier series (one sine and cosine component) and subtracted the modulation from the derived times. We additionally excluded from the fit any measurements with a residual larger than 0.25 cycles. 

We initially fitted each dataset (HIPPO and each of the ULTRACAM filters) independently to probe for dependence of the pulse arrival times with wavelength (as seen for AR~Sco \cite{Gaibor2020,Pelisoli2022c}). We found the ephemeris to be consistent between datasets, and hence fitted all data together. Following the described procedure, we obtained the following spin ephemeris:
\begin{eqnarray}
BJD(TDB) &=& 2459772.142522(24) + 0.0036961693(10)E,
\label{spin_eph}
\end{eqnarray}
where $E$ is an integral cycle number. The best fit and uncertainties were determined via bootstrapping.

Additionally, we have probed for the occurrence of spin-up or spin-down, as observed for AR~Sco\cite{Marsh2016,Stiller2018,Gaibor2020,Pelisoli2022c}, by fitting a quadratic ephemeris and performing an $F$-test to determine whether the addition of a quadratic term significantly improved the fit. We have found that for none of the our datasets a quadratic fit represented a significant improvement, which is unsurprising given our short baseline. We attempted to increase our baseline by deriving a pulse arrival measurement from the ATLAS $c$ data, which shows a hint of the spin period in its Fourier transform. Our approach was to subtract the orbital modulation modelled by a Fourier series, and fit the residuals with a cosine with period fixed to the spin period, but we found that the large uncertainties resulted in a poor fit with an amplitude consistent with zero. Therefore, a constraint on the spin period change could not be obtained with the current data.

\subsection*{The possible occurrence of flares}

The optical observations of \targ\ often showed hints of flaring, with the amplitude of the possible flares even dominating over the orbital modulation in a few occasions (see Supplementary Figure~\ref{fig:mookodi}). We studied the possibility of flares in more detail using the simultaneous \xmmn\ and ULTRACAM data, as shown in Figure~\ref{fig:flares}. The simultaneous data covers orbital phases 0.05 to 0.26, where typically the optical pulses are weak and less sharp. Yet, several narrow features consistent with the location of spin maxima are seen (in particular at spin cycles 18532--18535). The X-ray pulses, in contrast, are displaced in phase compared to narrow pulses, but coincide with wider features in the optical data. Our interpretation is that the short-lived features seen in the ULTRACAM data originate in the magnetosphere of the white dwarf, thus repeating on the spin period. As seen in Fig.~\ref{phaseFT}, they are associated with the narrow pulses seen in radio data. The broader features seen in both ULTRACAM and \xmmn\ show an offset compared to the narrow maxima (as also seen in Fig.~\ref{phaseFT}), suggesting they originate in another region in the system, possibly near the M-dwarf companion where the strongest emission is observed.

Associating the narrow features with the magnetosphere of white dwarf and the broader pulses with emission elsewhere, causing a phase delay, still leaves at least one unidentified feature around cycle 18534.5. The amplitude of this feature shows no apparent colour dependency (Extended Data Figure~\ref{fig:flare_colour}), unlike M-dwarf flares. We suggest that these are flares induced by accretion along the white dwarf's magnetic field lines. If this interpretation is correct, it would imply that, unlike AR~Sco, \targ\ is not completely detached yet, thus being in an earlier stage of evolution according to the model of Schreiber et al. 2021\cite{Schreiber2021}. Continuous monitoring of this source to probe the occurrence of flares will allow testing this hypothesis.

\subsection*{Constraining the white dwarf temperature}

The temperature of the white dwarf is a crucial parameter, especially in the context of the theoretical model proposed to explain the origin of binary white dwarf pulsars \cite{Schreiber2021}. In this model, the generation of the white dwarf magnetic field, which eventually becomes strong enough to connect with the M-dwarf magnetic field and provide a synchronising torque that explains the fast spin-down of systems like AR~Sco, has been attributed to crystallisation- and rotation-driven dynamo. Therefore, the white dwarf must have cooled down enough for crystallisation to progress in the core. For the range of masses derived, crystallisation will start below temperatures of $35\,000 - 14\,000$~K for a carbon-oxygen core white dwarf, depending both on the white dwarf mass and on the thickness of its atmosphere \cite{Bedard2020}. 

To constrain the white dwarf temperature, we first considered the observed GALEX magnitudes, where the white dwarf completely dominates over the companion.  We calculated the absolute magnitude given the {\it Gaia} distance and taking extinction at \targ's distance and sky direction \cite{Lallement2019} into account, and compared to synthetic magnitudes for cooling models \cite{Tremblay2011,Bedard2020}. For a mass of 1.0~M$_{\odot}$, the observed magnitudes are consistent with a maximum temperature of $\sim 15\,000$~K, above which the white dwarf flux would exceed the observed flux. For 1.2~M$_{\odot}$, the upper limit is $18\,000$~K. Another constrain can be obtained from the lack of any visible absorption lines from the white dwarf. To derive an upper limit based on that, our approach was to subtract a scaled white dwarf model from the observed spectrum and identify when that introduced a slope near the emission lines, suggesting that the white dwarf absorption line would be visible at that temperature. The models were scaled taking into account \targ's distance and a white dwarf radius interpolated from cooling models for each temperature\cite{Tremblay2011,Bedard2020}, assuming a mass of 1.0~M$_{\odot}$ for $\log~g = 8.5$ and 1.2~M$_{\odot}$ for $\log~g = 9.0$. We note that radius changes are minimal and such that the $\log~g$ change is negligible for fixed a mass. To numerically determine when there is a significant slope, we fitted the continuum near the H$\gamma$ line, where the spectrum is close to flat, after subtracting the scaled model. We fit both using a constant and a first order polynomial (indicating a slope), and perform an $F$-test to determine when the slope becomes significant. This suggested $T_1 < 13\,000$~K, above which the wings of the absorption lines of the white dwarf would be detectable (Supplementary Figure~\ref{wdteff}). For a mass of 1.2~M$_{\odot}$, the upper limit is less restrictive than the SED (23\,000~K).

For 1.2~M$_{\odot}$, crystallisation starts around 20\,000~K for a carbon-oxygen core\cite{Bedard2020}, and around 24\,000~K for a oxygen-neon core\cite{Camisassa2019}, therefore the derived upper limit suggestions crystallisation already started. The 1.0~M$_{\odot}$ case is less conclusive: crystallisation starts around our derived upper limit \cite{Bedard2020}. Hence the derive maximum temperature cannot rule out that the core is not significantly crystallised, but is consistent with it being at least starting to crystallise. Since the 1.0~M$_{\odot}$ case is the less conclusive in terms of crystallisation, it is the one we have focused on and illustrated in the main text.

\subsection*{Doppler tomography and the origin of emission}

The X-shooter spectra cover a full orbital period, allowing us to employ the Doppler tomography technique \cite{Marsh2001}, which maps the observed line profiles at different orbital phases into velocity space. This allows the emission distribution in the binary to be mapped. We computed Doppler maps for H$\beta$, H$\alpha$, and the CaII triplet around 8500~\AA. The trailed line profiles for these lines can be seen in Supplementary Figure~\ref{trails}. For H$\alpha$ (panel b), in additional to the sinusoidal profile due to orbital motion, satellite lines extending to higher velocities can be seen, similarly to what has been observed for AR~Sco \cite{Garnavich2019}. Frame-transfer spectroscopy also showed that H$\alpha$ is seen to pulsate with a period closer to the spin period (see Supplementary Figure~\ref{trail_alpha}). Corresponding Doppler maps are shown in Supplementary Figure~\ref{doppler}. It can be seen that the emission originates mainly at the irradiated face of the M-dwarf, with a somewhat extended region towards the white dwarf for H$\alpha$ and H$\beta$. H$\alpha$ additionally shows signs of prominences around -200 and 200 km/s, a pattern also seen in the Doppler maps of AR~Sco \cite{Garnavich2019, Pelisoli2022c}. Though these prominences could be attributed to accumulated material in these regions, possibly due to the magnetic field altering the Roche geometry and placing stable Lagrange points at these locations \cite{Schmidtobreick2012}, an alternative is that the H$\alpha$ line profile has components that are not kinematic in origin, but result from variations in optical depth or other changes in the radiative transfer in the M-dwarf photosphere that change the line profile independently of velocity. This suggestion is motivated by the fact that H$\alpha$ is seen to show atypical line profiles even in some detached binaries of M-dwarfs with white dwarf companions \cite{Maxted1998}.

\subsection*{Photopolarimetry and modelling of the emission}

An example of a single night's (29 June 2022) time-series photo-polarimetry data is shown in Supplementary Figure~\ref{egHIPPO}. The orbital modulation and the short period pulses are clearly seen in the photometry. The circular polarisation appears consistent with zero percent. The linear polarisation averages around 4 percent, although some single data points show larger values, up to 12 percent (see Supplementary Figure~\ref{egHIPPO2}, which zooms into the region where the linear polarisation pulses can be seen). The linear polarisation data is of insufficient signal-to-noise to be binned with enough time resolution to show the photometric pulses. Therefore, we subjected the entire series of photo-polarimetry to a Fourier analysis, for which the amplitude spectrum is presented in Supplementary Figure~\ref{FTHIPPO}. The amplitude spectra, for both the photometry and the linearly polarised flux, display the most prominent peaks at the orbital frequency ($\Omega$), the spin frequency ($\omega$) and twice the beat frequency ($2(\omega-\Omega)$). There were no significant frequencies in the amplitude spectrum of the circular polarisation.

To increase the signal-to-noise, we phase-binned and folded the data on the spin and beat frequencies as a function of orbital phase (Figure~\ref{fig:phot_pol}, left-hand panels). These so-called dynamic pulse profiles show how the beat and spin variations are modulated on the orbital period. Immediately obvious is that both the photometric and linear pulses (spin and beat) evolve in amplitude over the orbital cycle, peaking at $\sim 0.4$ in orbital phase. The spin and beat pulses have a diagonal structure as they appear to drift later and earlier, respectively, as a function of orbital phase, thus indicating that the polarised emission has both spin and beat components. In addition, there is a fainter photometric pulse during orbital phases $\sim 0.6-1.1$. We performed the same exercise assuming the dominant frequency to be that of the beat, rather than the spin, which resulted in smeared features (see Supplementary Figure~\ref{fig:wrong_freqs}). This significant increase in smearing when the beat is assumed to be dominant indicates that the data have not been folded on the correct spin and beat frequencies. This was our main motivation behind the interpretation of spin as the dominant frequency. In addition, the position angle of linear polarisation seems to be better defined when assuming the dominant peak to be the spin, though this is marginal at the moment and will require further follow-up to be confirmed (see Supplementary Figure~\ref{fig:polangle}).

The observed dynamic pulse profile is remarkably identical (in morphology and orbital phasing) to the main pulse in AR~Sco (see fig. 2 of Potter \& Buckley 2018\cite{PotterBuckley2018}). On closer inspection, there may be an indication that the linearly polarised spin and beat pulses are diagonally split, in the same manner as the main pulse in AR Sco, given that there seems to be a valley of lower intensity between peaks of higher intensity at each end of the profile. Future observations with higher signal-to-noise ratio will confirm whether this is a real feature.

Given the observed photopolarimetric similarities between \targ\ and AR Sco, we adapted the simple synchrotron model previously used in the literature to explain AR Sco's emission\cite{PotterBuckley2018}. The model photopolarimetric emission is shown as dynamic pulse profiles in the right-hand panels of Figure~\ref{fig:phot_pol}. The model assumes a synchrotron emission source is locked in the white dwarf rotating frame, which receives a further injection of electrons as the white dwarf's magnetic field sweeps past the secondary star on the beat frequency. The magnetic field of the rotating white dwarf accelerates the electrons to relativistic speeds, resulting in beamed synchrotron emission. We used the AR~Sco model, but with a smaller inclination of 37$^{\circ}$ and with a single emission region instead of two. As can be seen from the right-hand panels of Figure~\ref{fig:phot_pol}, the model visually reproduces the observed dynamic pulse profile quite well, in particular the orbital phasing of the pulses and their morphology. The absence of a second emission region and a less significant splitting of the linear pulses in \targ\ compared to AR~Sco is simply explained as an inclination effect, i.e. \targ\ is a lower inclination version of AR~Sco.

\section*{Data availability}
The TESS data used in this work are public and can be accessed via the Barbara A. Mikulski Archive for Space Telescopes (\href{https://mast.stsci.edu/}{https://mast.stsci.edu/}). Other data will be become public after the proprietary time expires, but can be made available upon reasonable request to the corresponding author.

\section*{Code availability}
Any of the custom data analysis scripts used in this work can be made available upon reasonable request to the corresponding author.

\bibliography{J1912}

\begin{thebibliography}{10}
\urlstyle{rm}
\expandafter\ifx\csname url\endcsname\relax
  \def\url#1{\texttt{#1}}\fi
\expandafter\ifx\csname urlprefix\endcsname\relax\def\urlprefix{URL }\fi
\expandafter\ifx\csname doiprefix\endcsname\relax\def\doiprefix{DOI: }\fi
\providecommand{\bibinfo}[2]{#2}
\providecommand{\eprint}[2][]{\url{#2}}

\bibitem{Stanway2018}
\bibinfo{author}{{Stanway}, E.~R.} et~al.
\newblock \bibinfo{journal}{\bibinfo{title}{{VLA radio observations of AR
  Scorpii}}}.
\newblock {\it {\JournalTitle{\aap}}} \textbf{\bibinfo{volume}{611}},
  \bibinfo{pages}{A66} (\bibinfo{year}{2018}).

\bibitem{Takata2018}
\bibinfo{author}{{Takata}, J.} et~al.
\newblock \bibinfo{journal}{\bibinfo{title}{{A Non-thermal Pulsed X-Ray
  Emission of AR Scorpii}}}.
\newblock {\it {\JournalTitle{\apj}}} \textbf{\bibinfo{volume}{853}},
  \bibinfo{pages}{106} (\bibinfo{year}{2018}).

\bibitem{Marsh2016}
\bibinfo{author}{{Marsh}, T.~R.} et~al.
\newblock \bibinfo{journal}{\bibinfo{title}{{A radio-pulsing white dwarf binary
  star}}}.
\newblock {\it {\JournalTitle{\nat}}} \textbf{\bibinfo{volume}{537}},
  \bibinfo{pages}{374--377} (\bibinfo{year}{2016}).

\bibitem{Katz2017}
\bibinfo{author}{{Katz}, J.~I.}
\newblock \bibinfo{journal}{\bibinfo{title}{{AR Sco: A Precessing White Dwarf
  Synchronar?}}}
\newblock {\it {\JournalTitle{\apj}}} \textbf{\bibinfo{volume}{835}},
  \bibinfo{pages}{150} (\bibinfo{year}{2017}).

\bibitem{Takata2017}
\bibinfo{author}{{Takata}, J.}, \bibinfo{author}{{Yang}, H.} \&
  \bibinfo{author}{{Cheng}, K.~S.}
\newblock \bibinfo{journal}{\bibinfo{title}{{A Model for AR Scorpii: Emission
  from Relativistic Electrons Trapped by Closed Magnetic Field Lines of
  Magnetic White Dwarfs}}}.
\newblock {\it {\JournalTitle{\apj}}} \textbf{\bibinfo{volume}{851}},
  \bibinfo{pages}{143} (\bibinfo{year}{2017}).

\bibitem{PotterBuckley2018}
\bibinfo{author}{{Potter}, S.~B.} \& \bibinfo{author}{{Buckley}, D. A.~H.}
\newblock \bibinfo{journal}{\bibinfo{title}{{Time series photopolarimetry and
  modelling of the white dwarf pulsar in AR Scorpii}}}.
\newblock {\it {\JournalTitle{\mnras}}} \textbf{\bibinfo{volume}{481}},
  \bibinfo{pages}{2384--2392} (\bibinfo{year}{2018}).

\bibitem{Plessis2022}
\bibinfo{author}{{du Plessis}, L.} et~al.
\newblock \bibinfo{journal}{\bibinfo{title}{{Probing the non-thermal emission
  geometry of AR Sco via optical phase-resolved polarimetry}}}.
\newblock {\it {\JournalTitle{\mnras}}} \textbf{\bibinfo{volume}{510}},
  \bibinfo{pages}{2998--3010} (\bibinfo{year}{2022}).

\bibitem{Geng2016}
\bibinfo{author}{{Geng}, J.-J.}, \bibinfo{author}{{Zhang}, B.} \&
  \bibinfo{author}{{Huang}, Y.-F.}
\newblock \bibinfo{journal}{\bibinfo{title}{{A Model of White Dwarf Pulsar AR
  Scorpii}}}.
\newblock {\it {\JournalTitle{\apjl}}} \textbf{\bibinfo{volume}{831}},
  \bibinfo{pages}{L10} (\bibinfo{year}{2016}).

\bibitem{Barnes2007}
\bibinfo{author}{{Barnes}, S.~A.}
\newblock \bibinfo{journal}{\bibinfo{title}{{Ages for Illustrative Field Stars
  Using Gyrochronology: Viability, Limitations, and Errors}}}.
\newblock {\it {\JournalTitle{\apj}}} \textbf{\bibinfo{volume}{669}},
  \bibinfo{pages}{1167--1189} (\bibinfo{year}{2007}).

\bibitem{Hermes2017}
\bibinfo{author}{{Hermes}, J.~J.} et~al.
\newblock \bibinfo{journal}{\bibinfo{title}{{White Dwarf Rotation as a Function
  of Mass and a Dichotomy of Mode Line Widths: Kepler Observations of 27
  Pulsating DA White Dwarfs through K2 Campaign 8}}}.
\newblock {\it {\JournalTitle{\apjs}}} \textbf{\bibinfo{volume}{232}},
  \bibinfo{pages}{23} (\bibinfo{year}{2017}).

\bibitem{Corsico2019}
\bibinfo{author}{{C{\'o}rsico}, A.~H.}, \bibinfo{author}{{Althaus}, L.~G.},
  \bibinfo{author}{{Miller Bertolami}, M.~M.} \& \bibinfo{author}{{Kepler},
  S.~O.}
\newblock \bibinfo{journal}{\bibinfo{title}{{Pulsating white dwarfs: new
  insights}}}.
\newblock {\it {\JournalTitle{\aapr}}} \textbf{\bibinfo{volume}{27}},
  \bibinfo{pages}{7} (\bibinfo{year}{2019}).

\bibitem{Patterson1994}
\bibinfo{author}{{Patterson}, J.}
\newblock \bibinfo{journal}{\bibinfo{title}{{The DQ Herculis Stars}}}.
\newblock {\it {\JournalTitle{\pasp}}} \textbf{\bibinfo{volume}{106}},
  \bibinfo{pages}{209} (\bibinfo{year}{1994}).

\bibitem{Lyutikov2020}
\bibinfo{author}{{Lyutikov}, M.} et~al.
\newblock \bibinfo{journal}{\bibinfo{title}{{Magnetospheric interaction in
  white dwarf binaries AR Sco and AE Aqr}}}.
\newblock {\it {\JournalTitle{arXiv e-prints}}}
  \bibinfo{pages}{arXiv:2004.11474} (\bibinfo{year}{2020}).

\bibitem{1979ApJ...232..259G}
\bibinfo{author}{{Ghosh}, P.} \& \bibinfo{author}{{Lamb}, F.~K.}
\newblock \bibinfo{journal}{\bibinfo{title}{{Accretion by rotating magnetic
  neutron stars. II. Radial and vertical structure of the transition zone in
  disk accretion.}}}
\newblock {\it {\JournalTitle{\apj}}} \textbf{\bibinfo{volume}{232}},
  \bibinfo{pages}{259--276} (\bibinfo{year}{1979}).

\bibitem{Pala2022}
\bibinfo{author}{{Pala}, A.~F.} et~al.
\newblock \bibinfo{journal}{\bibinfo{title}{{Constraining the evolution of
  cataclysmic variables via the masses and accretion rates of their underlying
  white dwarfs}}}.
\newblock {\it {\JournalTitle{\mnras}}} \textbf{\bibinfo{volume}{510}},
  \bibinfo{pages}{6110--6132} (\bibinfo{year}{2022}).

\bibitem{Wynn1995}
\bibinfo{author}{{Wynn}, G.~A.} \& \bibinfo{author}{{King}, A.~R.}
\newblock \bibinfo{journal}{\bibinfo{title}{{Diamagnetic accretion in
  intermediate polars - I. Blob orbits and spin evolution}}}.
\newblock {\it {\JournalTitle{\mnras}}} \textbf{\bibinfo{volume}{275}},
  \bibinfo{pages}{9--21} (\bibinfo{year}{1995}).

\bibitem{Schreiber2021}
\bibinfo{author}{{Schreiber}, M.~R.}, \bibinfo{author}{{Belloni}, D.},
  \bibinfo{author}{{G{\"a}nsicke}, B.~T.}, \bibinfo{author}{{Parsons}, S.~G.}
  \& \bibinfo{author}{{Zorotovic}, M.}
\newblock \bibinfo{journal}{\bibinfo{title}{{The origin and evolution of
  magnetic white dwarfs in close binary stars}}}.
\newblock {\it {\JournalTitle{Nature Astronomy}}} \textbf{\bibinfo{volume}{5}},
  \bibinfo{pages}{648--654} (\bibinfo{year}{2021}).

\bibitem{Isern2017}
\bibinfo{author}{{Isern}, J.}, \bibinfo{author}{{Garc{\'\i}a-Berro}, E.},
  \bibinfo{author}{{K{\"u}lebi}, B.} \& \bibinfo{author}{{Lor{\'e}n-Aguilar},
  P.}
\newblock \bibinfo{journal}{\bibinfo{title}{{A Common Origin of Magnetism from
  Planets to White Dwarfs}}}.
\newblock {\it {\JournalTitle{\apjl}}} \textbf{\bibinfo{volume}{836}},
  \bibinfo{pages}{L28} (\bibinfo{year}{2017}).

\bibitem{Ginzburg2022}
\bibinfo{author}{{Ginzburg}, S.}, \bibinfo{author}{{Fuller}, J.},
  \bibinfo{author}{{Kawka}, A.} \& \bibinfo{author}{{Caiazzo}, I.}
\newblock \bibinfo{journal}{\bibinfo{title}{{Slow convection and fast rotation
  in crystallization-driven white dwarf dynamos}}}.
\newblock {\it {\JournalTitle{\mnras}}} \textbf{\bibinfo{volume}{514}},
  \bibinfo{pages}{4111--4119} (\bibinfo{year}{2022}).

\bibitem{Liebert2015}
\bibinfo{author}{{Liebert}, J.}, \bibinfo{author}{{Ferrario}, L.},
  \bibinfo{author}{{Wickramasinghe}, D.~T.} \& \bibinfo{author}{{Smith}, P.~S.}
\newblock \bibinfo{journal}{\bibinfo{title}{{Enigmas from the Sloan Digital Sky
  Survey DR7 Kleinman White Dwarf Catalog}}}.
\newblock {\it {\JournalTitle{\apj}}} \textbf{\bibinfo{volume}{804}},
  \bibinfo{pages}{93} (\bibinfo{year}{2015}).

\bibitem{Parsons2021}
\bibinfo{author}{{Parsons}, S.~G.} et~al.
\newblock \bibinfo{journal}{\bibinfo{title}{{Magnetic white dwarfs in
  post-common-envelope binaries}}}.
\newblock {\it {\JournalTitle{\mnras}}} \textbf{\bibinfo{volume}{502}},
  \bibinfo{pages}{4305--4327} (\bibinfo{year}{2021}).

\bibitem{Pala2020}
\bibinfo{author}{{Pala}, A.~F.} et~al.
\newblock \bibinfo{journal}{\bibinfo{title}{{A Volume-limited Sample of
  Cataclysmic Variables from Gaia DR2: Space Density and Population
  Properties}}}.
\newblock {\it {\JournalTitle{\mnras}}} \textbf{\bibinfo{volume}{494}},
  \bibinfo{pages}{3799--3827} (\bibinfo{year}{2020}).

\bibitem{Kato2021a}
\bibinfo{author}{{Kato}, T.} \& \bibinfo{author}{{Kojiguchi}, N.}
\newblock \bibinfo{journal}{\bibinfo{title}{{ZTF J185139.81+171430.3 =
  ZTF18abnbzvx: the second white dwarf pulsar?}}}
\newblock {\it {\JournalTitle{arXiv e-prints}}}
  \bibinfo{pages}{arXiv:2107.09913} (\bibinfo{year}{2021}).

\bibitem{Kato2021b}
\bibinfo{author}{{Kato}, T.}, \bibinfo{author}{{Hambsch}, F.-J.},
  \bibinfo{author}{{Pavlenko}, E.~P.} \& \bibinfo{author}{{Sosnovskij}, A.~A.}
\newblock \bibinfo{journal}{\bibinfo{title}{{Orbital and spin periods of the
  candidate white dwarf pulsar ASASSN-V J205543.90+240033.5}}}.
\newblock {\it {\JournalTitle{arXiv e-prints}}}
  \bibinfo{pages}{arXiv:2109.03979} (\bibinfo{year}{2021}).

\bibitem{Kato2022}
\bibinfo{author}{{Kato}, T.}
\newblock \bibinfo{journal}{\bibinfo{title}{{Gaia22ayj: outburst from a deeply
  eclipsing 9.36-min binary?}}}
\newblock {\it {\JournalTitle{arXiv e-prints}}}
  \bibinfo{pages}{arXiv:2203.13975} (\bibinfo{year}{2022}).

\bibitem{Pelisoli23}
\bibinfo{author}{{Pelisoli}, I.} et~al.
\newblock \bibinfo{journal}{\bibinfo{title}{{A targeted search for binary white
  dwarf pulsars using {\it Gaia} and WISE}}}.
\newblock {\it {\JournalTitle{In prep.}}}  (\bibinfo{year}{2023}).

\bibitem{ultracam}
\bibinfo{author}{{Dhillon}, V.~S.} et~al.
\newblock \bibinfo{journal}{\bibinfo{title}{{ULTRACAM: an ultrafast,
  triple-beam CCD camera for high-speed astrophysics}}}.
\newblock {\it {\JournalTitle{\mnras}}} \textbf{\bibinfo{volume}{378}},
  \bibinfo{pages}{825--840} (\bibinfo{year}{2007}).

\bibitem{erosita}
\bibinfo{author}{{Predehl}, P.} et~al.
\newblock \bibinfo{journal}{\bibinfo{title}{{The eROSITA X-ray telescope on
  SRG}}}.
\newblock {\it {\JournalTitle{\aap}}} \textbf{\bibinfo{volume}{647}},
  \bibinfo{pages}{A1} (\bibinfo{year}{2021}).

\bibitem{Potter2010}
\bibinfo{author}{{Potter}, S.~B.} et~al.
\newblock \bibinfo{journal}{\bibinfo{title}{{Polarized QPOs from the INTEGRAL
  polar IGRJ14536-5522 (=Swift J1453.4-5524)}}}.
\newblock {\it {\JournalTitle{MNRAS}}} \textbf{\bibinfo{volume}{402}},
  \bibinfo{pages}{1161--1170} (\bibinfo{year}{2010}).

\bibitem{tess}
\bibinfo{author}{{Ricker}, G.~R.} et~al.
\newblock \bibinfo{journal}{\bibinfo{title}{{Transiting Exoplanet Survey
  Satellite (TESS)}}}.
\newblock {\it {\JournalTitle{Journal of Astronomical Telescopes, Instruments,
  and Systems}}} \textbf{\bibinfo{volume}{1}}, \bibinfo{pages}{014003}
  (\bibinfo{year}{2015}).

\bibitem{atlas}
\bibinfo{author}{{Tonry}, J.~L.} et~al.
\newblock \bibinfo{journal}{\bibinfo{title}{{ATLAS: A High-cadence All-sky
  Survey System}}}.
\newblock {\it {\JournalTitle{\pasp}}} \textbf{\bibinfo{volume}{130}},
  \bibinfo{pages}{064505} (\bibinfo{year}{2018}).

\bibitem{crts}
\bibinfo{author}{{Drake}, A.~J.} et~al.
\newblock \bibinfo{title}{{The Catalina Real-time Transient Survey}}.
\newblock In \bibinfo{editor}{{Griffin}, E.}, \bibinfo{editor}{{Hanisch}, R.}
  \& \bibinfo{editor}{{Seaman}, R.} (eds.) {\bibinfo{booktitle}{New Horizons in
  Time Domain Astronomy}}, vol. \bibinfo{volume}{285},
  \bibinfo{pages}{306--308} (\bibinfo{year}{2012}).

\bibitem{asassn}
\bibinfo{author}{{Shappee}, B.} et~al.
\newblock \bibinfo{title}{{All Sky Automated Survey for SuperNovae (ASAS-SN or
  ``Assassin'')}}.
\newblock In {\bibinfo{booktitle}{American Astronomical Society Meeting
  Abstracts \#223}}, vol. \bibinfo{volume}{223} of {\bibinfo{series}{American
  Astronomical Society Meeting Abstracts}}, \bibinfo{pages}{236.03}
  (\bibinfo{year}{2014}).

\bibitem{goodman}
\bibinfo{author}{{Clemens}, J.~C.}, \bibinfo{author}{{Crain}, J.~A.} \&
  \bibinfo{author}{{Anderson}, R.}
\newblock \bibinfo{title}{{The Goodman spectrograph}}.
\newblock In \bibinfo{editor}{{Moorwood}, A. F.~M.} \& \bibinfo{editor}{{Iye},
  M.} (eds.) {\bibinfo{booktitle}{Ground-based Instrumentation for Astronomy}},
  vol. \bibinfo{volume}{5492} of {\bibinfo{series}{Society of Photo-Optical
  Instrumentation Engineers (SPIE) Conference Series}},
  \bibinfo{pages}{331--340} (\bibinfo{year}{2004}).

\bibitem{Vernet2011}
\bibinfo{author}{{Vernet}, J.} et~al.
\newblock \bibinfo{journal}{\bibinfo{title}{{X-shooter, the new wide band
  intermediate resolution spectrograph at the ESO Very Large Telescope}}}.
\newblock {\it {\JournalTitle{\aap}}} \textbf{\bibinfo{volume}{536}},
  \bibinfo{pages}{A105} (\bibinfo{year}{2011}).

\bibitem{Caleb2022}
\bibinfo{author}{{Caleb}, M.} et~al.
\newblock \bibinfo{journal}{\bibinfo{title}{{Discovery of a radio-emitting
  neutron star with an ultra-long spin period of 76 s}}}.
\newblock {\it {\JournalTitle{Nature Astronomy}}} \textbf{\bibinfo{volume}{6}},
  \bibinfo{pages}{828--836} (\bibinfo{year}{2022}).

\bibitem{Eggleton1983}
\bibinfo{author}{{Eggleton}, P.~P.}
\newblock \bibinfo{journal}{\bibinfo{title}{{Aproximations to the radii of
  Roche lobes.}}}
\newblock {\it {\JournalTitle{\apj}}} \textbf{\bibinfo{volume}{268}},
  \bibinfo{pages}{368--369} (\bibinfo{year}{1983}).

\bibitem{Pelisoli2022c}
\bibinfo{author}{{Pelisoli}, I.} et~al.
\newblock \bibinfo{journal}{\bibinfo{title}{{Long-term photometric monitoring
  and spectroscopy of the white dwarf pulsar AR Scorpii}}}.
\newblock {\it {\JournalTitle{\mnras}}} \textbf{\bibinfo{volume}{516}},
  \bibinfo{pages}{5052--5066} (\bibinfo{year}{2022}).

\bibitem{Bailer-Jones2021}
\bibinfo{author}{{Bailer-Jones}, C.~A.~L.}, \bibinfo{author}{{Rybizki}, J.},
  \bibinfo{author}{{Fouesneau}, M.}, \bibinfo{author}{{Demleitner}, M.} \&
  \bibinfo{author}{{Andrae}, R.}
\newblock \bibinfo{journal}{\bibinfo{title}{{Estimating Distances from
  Parallaxes. V. Geometric and Photogeometric Distances to 1.47 Billion Stars
  in Gaia Early Data Release 3}}}.
\newblock {\it {\JournalTitle{\aj}}} \textbf{\bibinfo{volume}{161}},
  \bibinfo{pages}{147} (\bibinfo{year}{2021}).

\bibitem{Eason1992}
\bibinfo{author}{{Eason}, E. L.~E.}, \bibinfo{author}{{Giampapa}, M.~S.},
  \bibinfo{author}{{Radick}, R.~R.}, \bibinfo{author}{{Worden}, S.~P.} \&
  \bibinfo{author}{{Hege}, E.~K.}
\newblock \bibinfo{journal}{\bibinfo{title}{{Spectroscopic and Photometric
  Observations of a Five-Magnitude Flare Event on UV CETI}}}.
\newblock {\it {\JournalTitle{\aj}}} \textbf{\bibinfo{volume}{104}},
  \bibinfo{pages}{1161} (\bibinfo{year}{1992}).

\bibitem{Stepanov1995}
\bibinfo{author}{{Stepanov}, A.~V.} et~al.
\newblock \bibinfo{journal}{\bibinfo{title}{{Multifrequency observations of a
  flare on UV Ceti.}}}
\newblock {\it {\JournalTitle{\aap}}} \textbf{\bibinfo{volume}{299}},
  \bibinfo{pages}{739} (\bibinfo{year}{1995}).

\bibitem{juric2008}
\bibinfo{author}{{Juri{\'c}}, M.} et~al.
\newblock \bibinfo{journal}{\bibinfo{title}{{The Milky Way Tomography with
  SDSS. I. Stellar Number Density Distribution}}}.
\newblock {\it {\JournalTitle{\apj}}} \textbf{\bibinfo{volume}{673}},
  \bibinfo{pages}{864--914} (\bibinfo{year}{2008}).

\bibitem{Dhillon2021}
\bibinfo{author}{{Dhillon}, V.~S.} et~al.
\newblock \bibinfo{journal}{\bibinfo{title}{{HiPERCAM: a quintuple-beam,
  high-speed optical imager on the 10.4-m Gran Telescopio Canarias}}}.
\newblock {\it {\JournalTitle{\mnras}}} \textbf{\bibinfo{volume}{507}},
  \bibinfo{pages}{350--366} (\bibinfo{year}{2021}).

\bibitem{skymapper}
\bibinfo{author}{{Onken}, C.~A.} et~al.
\newblock \bibinfo{journal}{\bibinfo{title}{{SkyMapper Southern Survey: Second
  data release (DR2)}}}.
\newblock {\it {\JournalTitle{\pasa}}} \textbf{\bibinfo{volume}{36}},
  \bibinfo{pages}{e033} (\bibinfo{year}{2019}).

\bibitem{molecfit1}
\bibinfo{author}{{Smette}, A.} et~al.
\newblock \bibinfo{journal}{\bibinfo{title}{{Molecfit: A general tool for
  telluric absorption correction. I. Method and application to ESO
  instruments}}}.
\newblock {\it {\JournalTitle{\aap}}} \textbf{\bibinfo{volume}{576}},
  \bibinfo{pages}{A77} (\bibinfo{year}{2015}).

\bibitem{molecfit2}
\bibinfo{author}{{Kausch}, W.} et~al.
\newblock \bibinfo{journal}{\bibinfo{title}{{Molecfit: A general tool for
  telluric absorption correction. II. Quantitative evaluation on
  ESO-VLT/X-Shooterspectra}}}.
\newblock {\it {\JournalTitle{\aap}}} \textbf{\bibinfo{volume}{576}},
  \bibinfo{pages}{A78} (\bibinfo{year}{2015}).

\bibitem{Burgh2003}
\bibinfo{author}{{Burgh}, E.~B.} et~al.
\newblock \bibinfo{title}{{Prime Focus Imaging Spectrograph for the Southern
  African Large Telescope: optical design}}.
\newblock In \bibinfo{editor}{{Iye}, M.} \& \bibinfo{editor}{{Moorwood}, A.
  F.~M.} (eds.) {\bibinfo{booktitle}{Instrument Design and Performance for
  Optical/Infrared Ground-based Telescopes}}, vol. \bibinfo{volume}{4841} of
  {\bibinfo{series}{Society of Photo-Optical Instrumentation Engineers (SPIE)
  Conference Series}}, \bibinfo{pages}{1463--1471} (\bibinfo{year}{2003}).

\bibitem{Kobulnicky2003}
\bibinfo{author}{{Kobulnicky}, H.~A.} et~al.
\newblock \bibinfo{title}{{Prime focus imaging spectrograph for the Southern
  African large telescope: operational modes}}.
\newblock In \bibinfo{editor}{{Iye}, M.} \& \bibinfo{editor}{{Moorwood}, A.
  F.~M.} (eds.) {\bibinfo{booktitle}{Instrument Design and Performance for
  Optical/Infrared Ground-based Telescopes}}, vol. \bibinfo{volume}{4841} of
  {\bibinfo{series}{Society of Photo-Optical Instrumentation Engineers (SPIE)
  Conference Series}}, \bibinfo{pages}{1634--1644} (\bibinfo{year}{2003}).

\bibitem{Smith2006}
\bibinfo{author}{{Smith}, M.~P.} et~al.
\newblock \bibinfo{title}{{The prime focus imaging spectrograph for the
  Southern African Large Telescope: structural and mechanical design and
  commissioning}}.
\newblock In \bibinfo{editor}{{McLean}, I.~S.} \& \bibinfo{editor}{{Iye}, M.}
  (eds.) {\bibinfo{booktitle}{Society of Photo-Optical Instrumentation
  Engineers (SPIE) Conference Series}}, vol. \bibinfo{volume}{6269} of
  {\bibinfo{series}{Society of Photo-Optical Instrumentation Engineers (SPIE)
  Conference Series}}, \bibinfo{pages}{62692A} (\bibinfo{year}{2006}).

\bibitem{Buckley2006}
\bibinfo{author}{{Buckley}, D. A.~H.}, \bibinfo{author}{{Swart}, G.~P.} \&
  \bibinfo{author}{{Meiring}, J.~G.}
\newblock \bibinfo{title}{{Completion and commissioning of the Southern African
  Large Telescope}}.
\newblock In \bibinfo{editor}{{Stepp}, L.~M.} (ed.)
  {\bibinfo{booktitle}{Society of Photo-Optical Instrumentation Engineers
  (SPIE) Conference Series}}, vol. \bibinfo{volume}{6267} of
  {\bibinfo{series}{Society of Photo-Optical Instrumentation Engineers (SPIE)
  Conference Series}}, \bibinfo{pages}{62670Z} (\bibinfo{year}{2006}).

\bibitem{Crawford2010}
\bibinfo{author}{{Crawford}, S.~M.} et~al.
\newblock \bibinfo{title}{{PySALT: the SALT science pipeline}}.
\newblock In \bibinfo{editor}{{Silva}, D.~R.}, \bibinfo{editor}{{Peck}, A.~B.}
  \& \bibinfo{editor}{{Soifer}, B.~T.} (eds.) {\bibinfo{booktitle}{Observatory
  Operations: Strategies, Processes, and Systems III}}, vol.
  \bibinfo{volume}{7737} of {\bibinfo{series}{Society of Photo-Optical
  Instrumentation Engineers (SPIE) Conference Series}}, \bibinfo{pages}{773725}
  (\bibinfo{year}{2010}).

\bibitem{Sunyaev2021}
\bibinfo{author}{{Sunyaev}, R.} et~al.
\newblock \bibinfo{journal}{\bibinfo{title}{{SRG X-ray orbital observatory. Its
  telescopes and first scientific results}}}.
\newblock {\it {\JournalTitle{\aap}}} \textbf{\bibinfo{volume}{656}},
  \bibinfo{pages}{A132} (\bibinfo{year}{2021}).

\bibitem{brunner+22}
\bibinfo{author}{{Brunner}, H.} et~al.
\newblock \bibinfo{journal}{\bibinfo{title}{{The eROSITA Final Equatorial Depth
  Survey (eFEDS). X-ray catalogue}}}.
\newblock {\it {\JournalTitle{\aap}}} \textbf{\bibinfo{volume}{661}},
  \bibinfo{pages}{A1} (\bibinfo{year}{2022}).

\bibitem{arnaud+96}
\bibinfo{author}{{Arnaud}, K.~A.}
\newblock \bibinfo{title}{{XSPEC: The First Ten Years}}.
\newblock In \bibinfo{editor}{{Jacoby}, G.~H.} \& \bibinfo{editor}{{Barnes},
  J.} (eds.) {\bibinfo{booktitle}{Astronomical Data Analysis Software and
  Systems V}}, vol. \bibinfo{volume}{101} of {\bibinfo{series}{Astronomical
  Society of the Pacific Conference Series}}, \bibinfo{pages}{17}
  (\bibinfo{year}{1996}).

\bibitem{Schwope23}
\bibinfo{author}{{Schwope}, I.} et~al.
\newblock \bibinfo{journal}{\bibinfo{title}{{X-ray properties of the white
  dwarf pulsar eRASSU J191213.9-441044}}}.
\newblock {\it {\JournalTitle{\aap}}} \textbf{\bibinfo{volume}{674}},
  \bibinfo{pages}{L9} (\bibinfo{year}{2023}).

\bibitem{casa2022}
\bibinfo{author}{{CASA Team}} et~al.
\newblock \bibinfo{journal}{\bibinfo{title}{{CASA, the Common Astronomy
  Software Applications for Radio Astronomy}}}.
\newblock {\it {\JournalTitle{\pasp}}} \textbf{\bibinfo{volume}{134}},
  \bibinfo{pages}{114501} (\bibinfo{year}{2022}).

\bibitem{hugo2022}
\bibinfo{author}{{Hugo}, B.~V.}, \bibinfo{author}{{Perkins}, S.},
  \bibinfo{author}{{Merry}, B.}, \bibinfo{author}{{Mauch}, T.} \&
  \bibinfo{author}{{Smirnov}, O.~M.}
\newblock \bibinfo{title}{{Tricolour: An Optimized SumThreshold Flagger for
  MeerKAT}}.
\newblock In \bibinfo{editor}{{Ruiz}, J.~E.}, \bibinfo{editor}{{Pierfedereci},
  F.} \& \bibinfo{editor}{{Teuben}, P.} (eds.)
  {\bibinfo{booktitle}{Astronomical Society of the Pacific Conference Series}},
  vol. \bibinfo{volume}{532} of {\bibinfo{series}{Astronomical Society of the
  Pacific Conference Series}}, \bibinfo{pages}{541} (\bibinfo{year}{2022}).

\bibitem{offringa2014}
\bibinfo{author}{{Offringa}, A.~R.} et~al.
\newblock \bibinfo{journal}{\bibinfo{title}{{WSCLEAN: an implementation of a
  fast, generic wide-field imager for radio astronomy}}}.
\newblock {\it {\JournalTitle{\mnras}}} \textbf{\bibinfo{volume}{444}},
  \bibinfo{pages}{606--619} (\bibinfo{year}{2014}).

\bibitem{heywood2020}
\bibinfo{author}{{Heywood}, I.}
\newblock \bibinfo{title}{{oxkat: Semi-automated imaging of MeerKAT
  observations}} (\bibinfo{year}{2020}).

\bibitem{Parsons2018}
\bibinfo{author}{{Parsons}, S.~G.} et~al.
\newblock \bibinfo{journal}{\bibinfo{title}{{The scatter of the M dwarf
  mass-radius relationship}}}.
\newblock {\it {\JournalTitle{\mnras}}} \textbf{\bibinfo{volume}{481}},
  \bibinfo{pages}{1083--1096} (\bibinfo{year}{2018}).

\bibitem{Littlefield2017}
\bibinfo{author}{{Littlefield}, C.} et~al.
\newblock \bibinfo{journal}{\bibinfo{title}{{Long-term Photometric Variations
  in the Candidate White-dwarf Pulsar AR Scorpii from K2, CRTS, and ASAS-SN
  Observations}}}.
\newblock {\it {\JournalTitle{\apjl}}} \textbf{\bibinfo{volume}{845}},
  \bibinfo{pages}{L7} (\bibinfo{year}{2017}).

\bibitem{Gaensicke2005}
\bibinfo{author}{{G{\"a}nsicke}, B.~T.} et~al.
\newblock \bibinfo{journal}{\bibinfo{title}{{Cataclysmic variables from a
  ROSAT/2MASS selection - I. Four new intermediate polars}}}.
\newblock {\it {\JournalTitle{\mnras}}} \textbf{\bibinfo{volume}{361}},
  \bibinfo{pages}{141--154} (\bibinfo{year}{2005}).

\bibitem{Brown2022}
\bibinfo{author}{{Brown}, A.~J.} et~al.
\newblock \bibinfo{journal}{\bibinfo{title}{{Characterizing eclipsing white
  dwarf M dwarf binaries from multiband eclipse photometry}}}.
\newblock {\it {\JournalTitle{\mnras}}} \textbf{\bibinfo{volume}{513}},
  \bibinfo{pages}{3050--3064} (\bibinfo{year}{2022}).

\bibitem{Kesseli2018}
\bibinfo{author}{{Kesseli}, A.~Y.}, \bibinfo{author}{{Muirhead}, P.~S.},
  \bibinfo{author}{{Mann}, A.~W.} \& \bibinfo{author}{{Mace}, G.}
\newblock \bibinfo{journal}{\bibinfo{title}{{Magnetic Inflation and Stellar
  Mass. II. On the Radii of Single, Rapidly Rotating, Fully Convective M-Dwarf
  Stars}}}.
\newblock {\it {\JournalTitle{\aj}}} \textbf{\bibinfo{volume}{155}},
  \bibinfo{pages}{225} (\bibinfo{year}{2018}).

\bibitem{Hauschildt1999}
\bibinfo{author}{{Hauschildt}, P.~H.}, \bibinfo{author}{{Allard}, F.} \&
  \bibinfo{author}{{Baron}, E.}
\newblock \bibinfo{journal}{\bibinfo{title}{{The NextGen Model Atmosphere Grid
  for 3000<=T$_{eff}$<=10,000 K}}}.
\newblock {\it {\JournalTitle{\apj}}} \textbf{\bibinfo{volume}{512}},
  \bibinfo{pages}{377--385} (\bibinfo{year}{1999}).

\bibitem{Knigge2011}
\bibinfo{author}{{Knigge}, C.}, \bibinfo{author}{{Baraffe}, I.} \&
  \bibinfo{author}{{Patterson}, J.}
\newblock \bibinfo{journal}{\bibinfo{title}{{The Evolution of Cataclysmic
  Variables as Revealed by Their Donor Stars}}}.
\newblock {\it {\JournalTitle{\apjs}}} \textbf{\bibinfo{volume}{194}},
  \bibinfo{pages}{28} (\bibinfo{year}{2011}).

\bibitem{Stiller2018}
\bibinfo{author}{{Stiller}, R.~A.} et~al.
\newblock \bibinfo{journal}{\bibinfo{title}{{High-time-resolution Photometry of
  AR Scorpii: Confirmation of the White Dwarf{\textquoteright}s Spin-down}}}.
\newblock {\it {\JournalTitle{\aj}}} \textbf{\bibinfo{volume}{156}},
  \bibinfo{pages}{150} (\bibinfo{year}{2018}).

\bibitem{Gaibor2020}
\bibinfo{author}{{Gaibor}, Y.}, \bibinfo{author}{{Garnavich}, P.~M.},
  \bibinfo{author}{{Littlefield}, C.}, \bibinfo{author}{{Potter}, S.~B.} \&
  \bibinfo{author}{{Buckley}, D.~A.~H.}
\newblock \bibinfo{journal}{\bibinfo{title}{{An improved spin-down rate for the
  proposed white dwarf pulsar AR scorpii}}}.
\newblock {\it {\JournalTitle{\mnras}}} \textbf{\bibinfo{volume}{496}},
  \bibinfo{pages}{4849--4856} (\bibinfo{year}{2020}).

\bibitem{Bedard2020}
\bibinfo{author}{{B{\'e}dard}, A.}, \bibinfo{author}{{Bergeron}, P.},
  \bibinfo{author}{{Brassard}, P.} \& \bibinfo{author}{{Fontaine}, G.}
\newblock \bibinfo{journal}{\bibinfo{title}{{On the Spectral Evolution of Hot
  White Dwarf Stars. I. A Detailed Model Atmosphere Analysis of Hot White
  Dwarfs from SDSS DR12}}}.
\newblock {\it {\JournalTitle{\apj}}} \textbf{\bibinfo{volume}{901}},
  \bibinfo{pages}{93} (\bibinfo{year}{2020}).

\bibitem{Lallement2019}
\bibinfo{author}{{Lallement}, R.} et~al.
\newblock \bibinfo{journal}{\bibinfo{title}{{Gaia-2MASS 3D maps of Galactic
  interstellar dust within 3 kpc}}}.
\newblock {\it {\JournalTitle{\aap}}} \textbf{\bibinfo{volume}{625}},
  \bibinfo{pages}{A135} (\bibinfo{year}{2019}).

\bibitem{Tremblay2011}
\bibinfo{author}{{Tremblay}, P.~E.}, \bibinfo{author}{{Bergeron}, P.} \&
  \bibinfo{author}{{Gianninas}, A.}
\newblock \bibinfo{journal}{\bibinfo{title}{{An Improved Spectroscopic Analysis
  of DA White Dwarfs from the Sloan Digital Sky Survey Data Release 4}}}.
\newblock {\it {\JournalTitle{\apj}}} \textbf{\bibinfo{volume}{730}},
  \bibinfo{pages}{128} (\bibinfo{year}{2011}).

\bibitem{Camisassa2019}
\bibinfo{author}{{Camisassa}, M.~E.} et~al.
\newblock \bibinfo{journal}{\bibinfo{title}{{VizieR Online Data Catalog:
  Ultra-massive white dwarfs evolution models (Camisassa+, 2019)}}}.
\newblock {\it {\JournalTitle{VizieR Online Data Catalog}}}
  \bibinfo{pages}{J/A+A/625/A87} (\bibinfo{year}{2019}).

\bibitem{Marsh2001}
\bibinfo{author}{{Marsh}, T.~R.}
\newblock \bibinfo{title}{{Doppler Tomography}}.
\newblock In \bibinfo{editor}{{Boffin}, H.~M.~J.}, \bibinfo{editor}{{Steeghs},
  D.} \& \bibinfo{editor}{{Cuypers}, J.} (eds.)
  {\bibinfo{booktitle}{Astrotomography, Indirect Imaging Methods in
  Observational Astronomy}}, vol. \bibinfo{volume}{573}, \bibinfo{pages}{1}
  (\bibinfo{year}{2001}).

\bibitem{Garnavich2019}
\bibinfo{author}{{Garnavich}, P.} et~al.
\newblock \bibinfo{journal}{\bibinfo{title}{{Driving the Beat: Time-resolved
  Spectra of the White Dwarf Pulsar AR Scorpii}}}.
\newblock {\it {\JournalTitle{\apj}}} \textbf{\bibinfo{volume}{872}},
  \bibinfo{pages}{67} (\bibinfo{year}{2019}).

\bibitem{Schmidtobreick2012}
\bibinfo{author}{{Schmidtobreick}, L.} et~al.
\newblock \bibinfo{journal}{\bibinfo{title}{{Discovery of
  H{\ensuremath{\alpha}} satellite emission in a low state of the SW Sextantis
  star BB Doradus}}}.
\newblock {\it {\JournalTitle{\mnras}}} \textbf{\bibinfo{volume}{422}},
  \bibinfo{pages}{731--737} (\bibinfo{year}{2012}).

\bibitem{Maxted1998}
\bibinfo{author}{{Maxted}, P.~F.~L.}, \bibinfo{author}{{Marsh}, T.~R.},
  \bibinfo{author}{{Moran}, C.}, \bibinfo{author}{{Dhillon}, V.~S.} \&
  \bibinfo{author}{{Hilditch}, R.~W.}
\newblock \bibinfo{journal}{\bibinfo{title}{{The mass and radius of the M dwarf
  companion to GD448}}}.
\newblock {\it {\JournalTitle{\mnras}}} \textbf{\bibinfo{volume}{300}},
  \bibinfo{pages}{1225--1232} (\bibinfo{year}{1998}).

\bibitem{astropy:2013}
\bibinfo{author}{{Astropy Collaboration}} et~al.
\newblock \bibinfo{journal}{\bibinfo{title}{{Astropy: A community Python
  package for astronomy}}}.
\newblock {\it {\JournalTitle{\aap}}} \textbf{\bibinfo{volume}{558}},
  \bibinfo{pages}{A33} (\bibinfo{year}{2013}).

\bibitem{astropy:2018}
\bibinfo{author}{{Price-Whelan}, A.~M.} et~al.
\newblock \bibinfo{journal}{\bibinfo{title}{{The Astropy Project: Building an
  Open-science Project and Status of the v2.0 Core Package}}}.
\newblock {\it {\JournalTitle{\aj}}} \textbf{\bibinfo{volume}{156}},
  \bibinfo{pages}{123} (\bibinfo{year}{2018}).

\end{thebibliography}

Correspondence and requests for materials should be addressed to Dr Ingrid Pelisoli (ingrid.pelisoli@warwick.ac.uk).

\section*{Acknowledgements} 

IP and TRM acknowledge funding by the UK's Science and Technology Facilities Council (STFC), grant ST/T000406/1. IP also acknowledges funding from a Warwick Astrophysics prize post-doctoral fellowship, made possible thanks to a generous philanthropic donation. IP was additionally supported in part by the National Science Foundation under Grant No. NSF PHY-1748958, and thanks the organisers of the KITP Program "White Dwarfs as Probes of the Evolution of Planets, Stars, the Milky Way and the Expanding Universe". SGP acknowledges the support of a STFC Ernest Rutherford Fellowship. VSD and ULTRACAM are funded by the UK’s Science and Technology Facilities Council (STFC), grant ST/V000853/1.
     
This research made extensive use of Astropy (http://www.astropy.org) a community-developed core Python package for Astronomy \cite{astropy:2013, astropy:2018}
     
This paper includes data collected by the TESS mission. Funding for the TESS mission is provided by the NASA Explorer Program.

Based in part on observations obtained at the Southern Astrophysical Research (SOAR) telescope, which is a joint project of the Ministério da Ciência, Tecnologia e Inovações do Brasil (MCTI/LNA), the US National Science Foundation’s NOIRLab, the University of North Carolina at Chapel Hill (UNC), and Michigan State University (MSU).

Based on observations collected at the European Organisation for Astronomical Research in the Southern Hemisphere under ESO programmes 109.24EM and 109.234F.

The SALT observations were obtained under the SALT Large Science Programme on transients (2021-2-LSP-001; PI: DAHB). Polish participation in SALT is funded by grant No. MEiN nr 2021/WK/01. DAHB and SBP acknowledge researh support by the National Research Foundation.

This work has made use of data from the European Space Agency (ESA) mission
{\it Gaia} (\url{https://www.cosmos.esa.int/gaia}), processed by the {\it Gaia}
Data Processing and Analysis Consortium (DPAC,
\url{https://www.cosmos.esa.int/web/gaia/dpac/consortium}). Funding for the DPAC
has been provided by national institutions, in particular the institutions
participating in the {\it Gaia} Multilateral Agreement.

The MeerKAT telescope is operated by the South African Radio Astronomy Observatory (SARAO), which is a facility of the National Research Foundation, an agency of the Department of Science and Innovation. The authors thank SARAO for the award of the MeerKAT Director's Discretionary Time.

Based on observations obtained with XMM-Newton, an ESA science mission with instruments and contributions directly funded by ESA Member States and NASA.

This work is based on data from eROSITA, the soft X-ray instrument aboard SRG, a joint Russian-German science mission supported by the Russian Space Agency (Roskosmos), in the interests of the Russian Academy of Sciences represented by its Space Research Institute (IKI), and the Deutsches Zentrum für Luft- und Raumfahrt (DLR). The SRG spacecraft was built by Lavochkin Association (NPOL) and its subcontractors, and is operated by NPOL with support from the Max Planck Institute for Extraterrestrial Physics (MPE). The development and construction of the eROSITA X-ray instrument was led by MPE, with contributions from the Dr. Karl Remeis Observatory Bamberg \& ECAP (FAU Erlangen-Nuernberg), the University of Hamburg Observatory, the Leibniz Institute for Astrophysics Potsdam (AIP), and the Institute for Astronomy and Astrophysics of the University of Tübingen, with the support of DLR and the Max Planck Society. The Argelander Institute for Astronomy of the University of Bonn and the Ludwig Maximilians Universität Munich also participated in the science preparation for eROSITA. The eROSITA data shown here were processed using the eSASS/NRTA software system developed by the German eROSITA consortium.

Support of the Deutsche Forschungsgemeinschaft (DFG) under grant number 536/37-1 is gratefully acknowledged.

\section*{Author contributions} 

All authors contributed to the work presented in this paper. IP wrote the manuscript and lead the follow-up and analysis of the system, with significant input from TRM. DAHB carried out follow-up observations at SAAO and contributed to analysis of the optical data. IH carried out reduction and analysis of the radio data. SBP carried out follow-up observations at SAAO and the analysis and modelling of polarimetric data. A. Schwope and A. Standke obtained and analysed the X-ray data. PAW obtained the radio data. SGP and MJG contributed to initial identification and analysis of the system. SOK, JM, and ADR contributed to observational follow-up. EB, AJB, VSD, MJD, PK, SPL, DIS, and JFW contributed to the maintenance of operations of ULTRACAM.

\section*{Competing interests statement}
The authors declare no competing interests.

\begin{extfigure*}[h!]
\centering
\includegraphics[width=0.8\hsize]{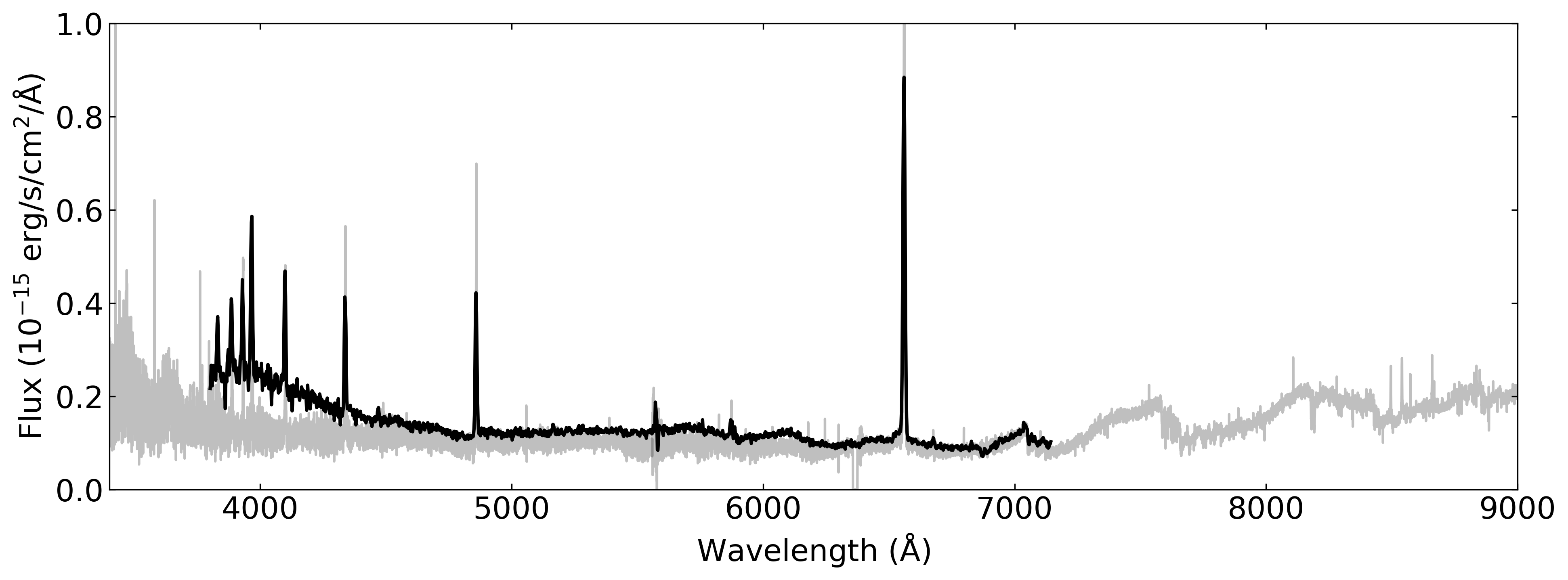}
\caption{{\bf Optical spectrum of \targ.} The black line shows the SOAR spectrum obtained for \targ, which confirmed its spectral characteristics to be similar to AR~Sco. The grey line shows X-Shooter spectra from the UVB and VIS arms obtained around the same orbital phase as the SOAR spectrum (0.85). The flux calibration of the SOAR spectrum is poor towards the blue due to reduced sensitivity.}
\label{fig:spec}
\end{extfigure*}

\begin{extfigure*}[h!]
\centering
\includegraphics[width=0.8\hsize]{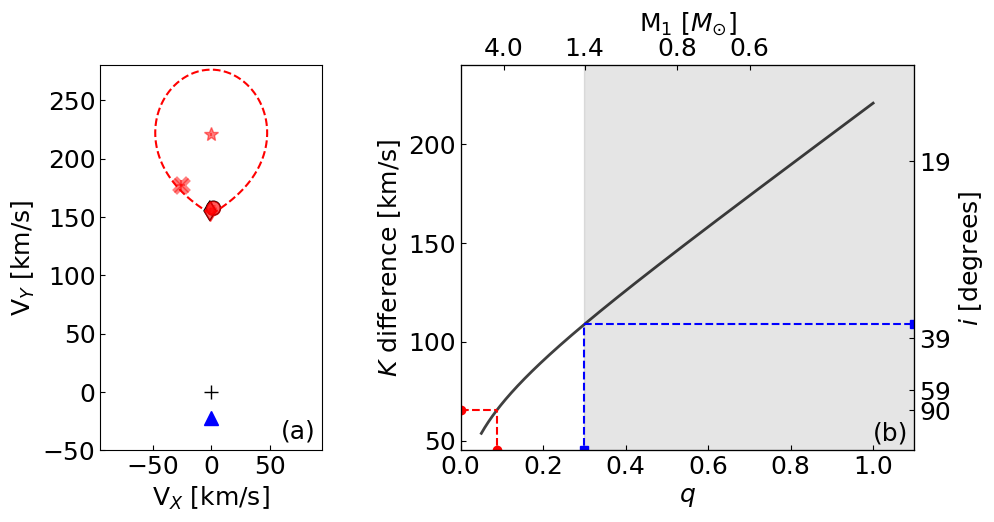}
\caption{{\bf Constraints from Roche geometry.} In panel (a) the star marks measurements from the NaII line (by our definition the centre of mass of the M-dwarf). The cross, diamond, and circle mark respectively H$\alpha$, H$\beta$, and H$\gamma$, which were fitted as $V = \gamma - V_X \cos(2\pi \phi)$  $+ V_Y \sin(2\pi \phi)$, where $\phi$ is the orbital phase and $\gamma$, the systemic velocity, was kept fixed to the previously determined value. H$\beta$, and H$\gamma$ give consistent measurements, whereas H$\alpha$ seems to trail the leading face of the M-dwarf. The red dashed line is the Roche lobe of the M-dwarf for $q=0.1$. The black cross and blue triangle mark the centre of mass of the system and of the white dwarf, respectively. The black line in panel (b) shows the expected difference between the irradiated face and centre of mass radial velocity semi-amplitudes as a function of $q$, assuming the M-dwarf fills its Roche lobe. The right-hand $y$-axis shows the required inclination to explain the detected $K$ difference. The observed value of $K$ difference sets a minimum for $q$, which would happen if the system was seen at 90$^{\circ}$ inclination, as indicated by the red dashed line. The area shaded in grey corresponds to $M_1$ values consistent with a white dwarf for a Roche lobe-filling companion, and implies the minimum $q$ value of 0.3. This minimum $q$ corresponds to a maximum inclination of 37$^{\circ}$ (blue dashed lines).}
\label{fig:roche}
\end{extfigure*}

\begin{extfigure*}[h!]
\centering
\includegraphics[width=0.7\hsize]{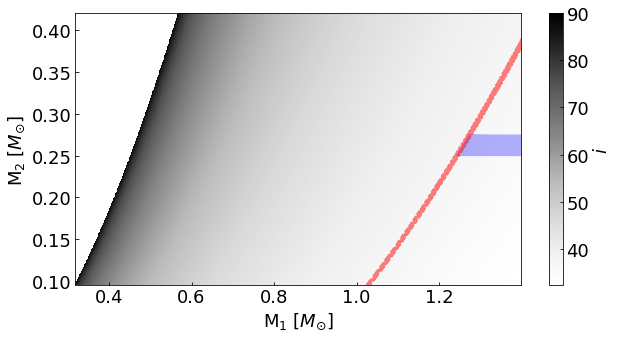}
\caption{{\bf Constraints from the binary mass function.} The colour map shows the required inclination to explain the observed $K_2$ for given values of $M_1$ and $M_2$ shown in the x- and y-axis. The red line marks the maximum inclination of 37$^{\circ}$, derived from Roche geometry, and the blue shaded area indicates the $M_2$ mass derived from a mass-radius relationship. Given the high systematic uncertainty on $M_2$, we adopt less strict constraints of $M_1 = 1.2\pm0.2$~M$_{\sun}$ and $M_2 = 0.25\pm0.05$~M$_{\sun}$.}
\label{fig:masses}
\end{extfigure*}

\begin{extfigure*}[h!]
\centering
\includegraphics[width=0.5\hsize]{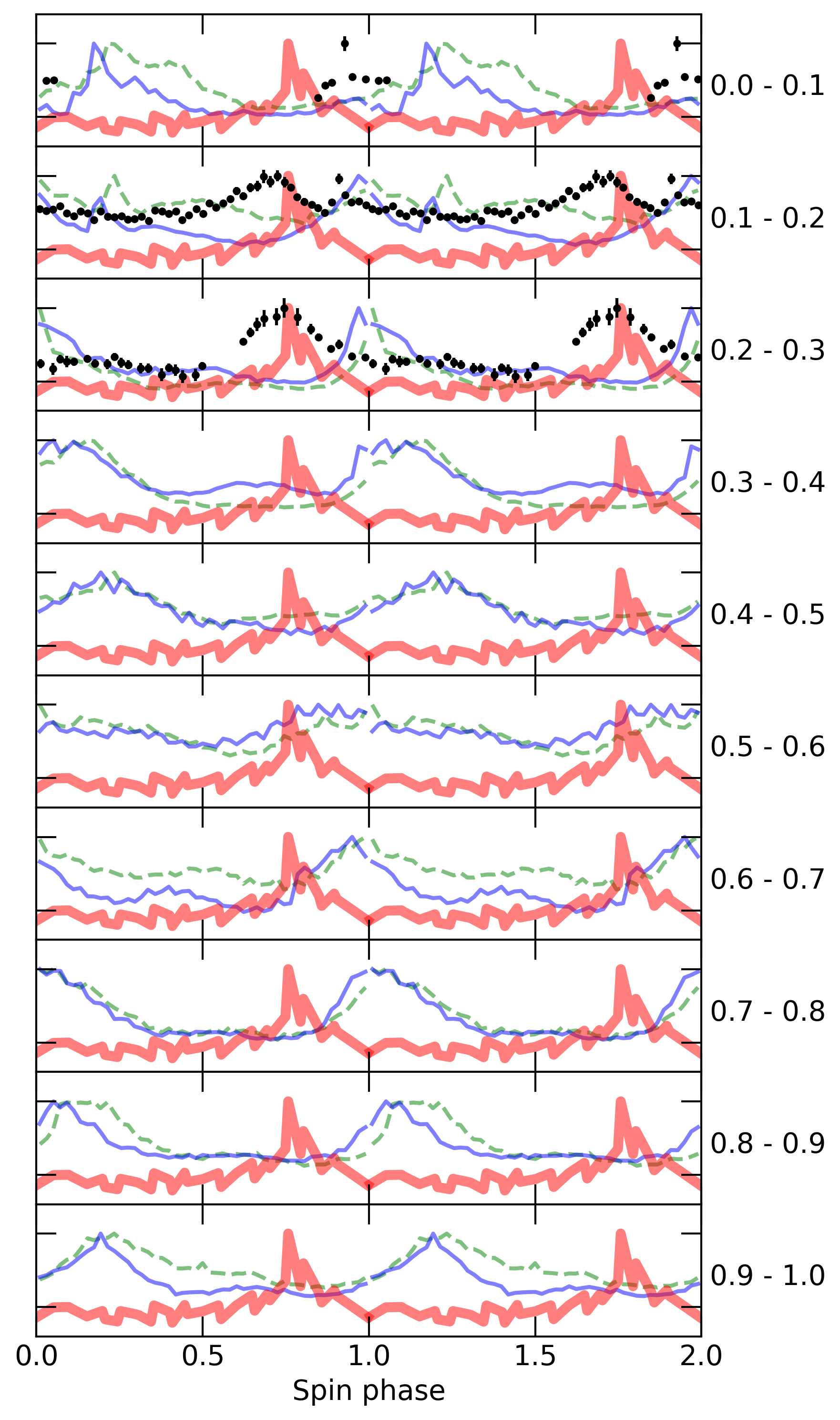}
\caption{{\bf Pulse shape for different orbital phases and nights.} The thick red line shows all the X-ray data averaged and folded on the spin ephemeris. The thin lines and symbols show ULTRACAM $g_s$ data averaged and folded on the same ephemeris, but considering data only within the orbital phase ranges shown on the right of the plot. The green dashed line shows data taken on 2022 June 07, the black symbols are data taken on 2022 September 17 (simultaneously with the X-ray data), and the solid blue line shows data for 2022 September 23. All data were normalised to the strongest peak to facilitate comparison. As also seen in Fig.~\ref{phaseFT}, the peak of the X-ray pulses does not align with the bulk of optical pulses. However, it does align with the optical peaks observed on 2022 September 17. This difference cannot be attributed to uncertainty in the ephemeris, given the agreement between data taken on nights before and after the X-ray observations. Additional simultaneous data is needed to determine the cause of misalignment, which could possibly be due to sporadic changes on pulse profile. }
\label{fig:pulses}
\end{extfigure*}

\begin{extfigure*}[h!]
\centering
\includegraphics[width=0.5\hsize]{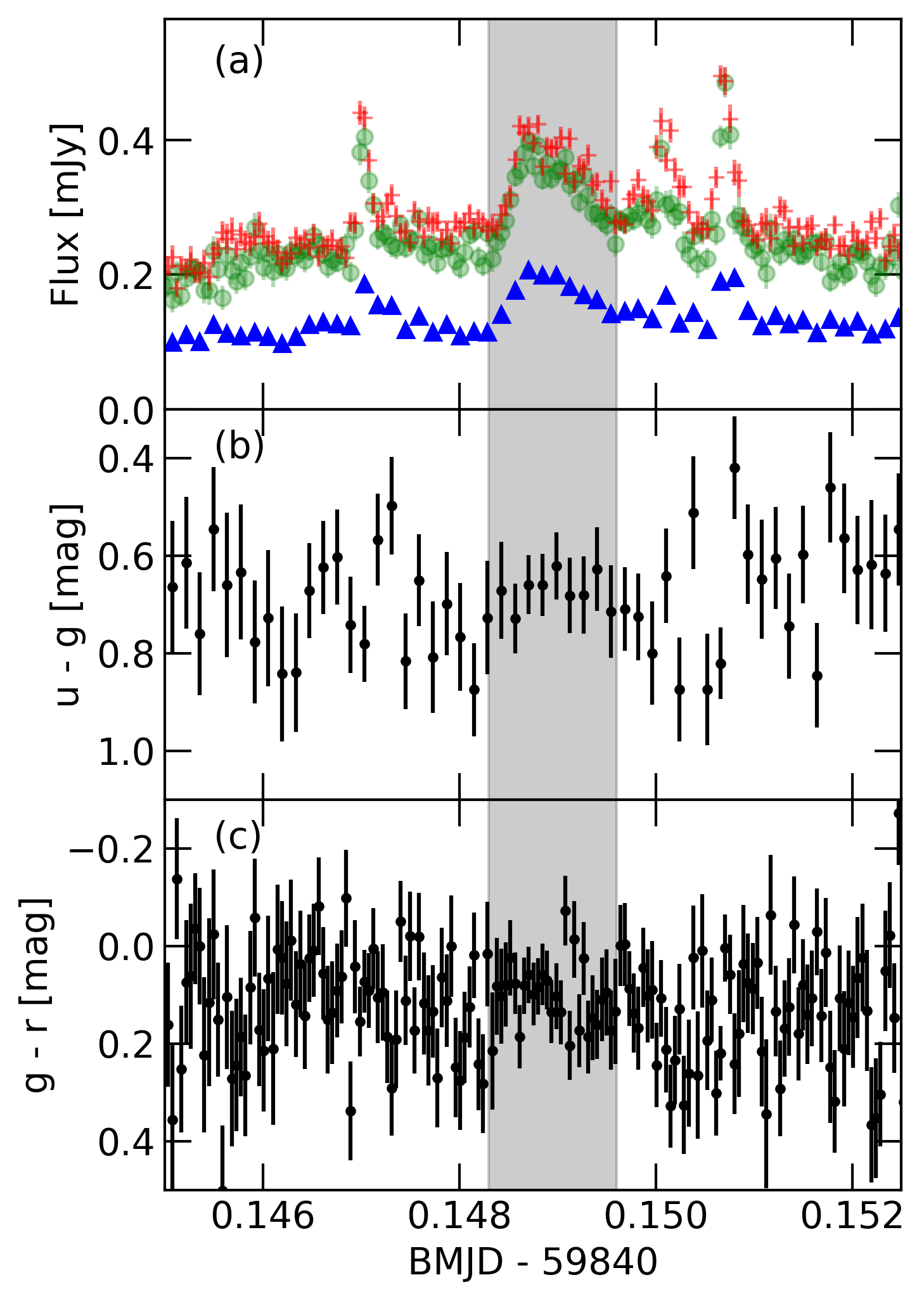}
\caption{{\bf Flux and colour of the possible flare.} Panel (a) shows the flux in the $u_s$ (blue triangles), $g_s$ (green circles), and $r_s$ (red crosses) bands in the region of the feature that we identify as a flare (marked by the shaded grey area). Panels (b) and (c) show the $u_s-g_s$ and $g_s-r_s$ colours. Unlike typical M-dwarf flares, there is no evidence of flux increase towards the blue.}
\label{fig:flare_colour}
\end{extfigure*}

\clearpage

\section*{Supplementary Information}

\begin{suppfigure*}[h!]
\centering
\includegraphics[width=0.8\hsize]{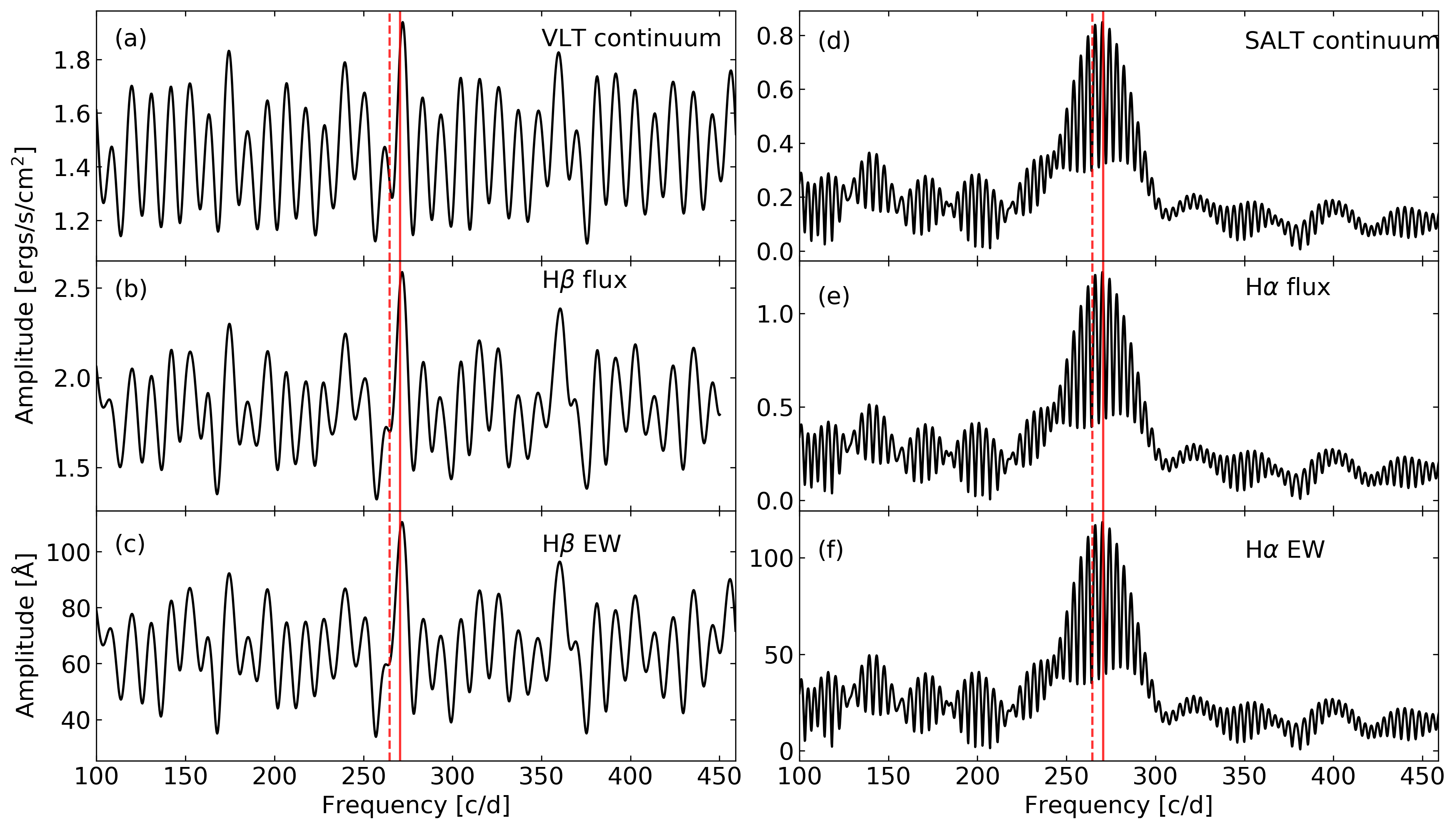}
\caption{{\bf Fourier transforms of data derived from spectroscopy.} Panels (a)-(c) show Fourier transforms of the continuum, H$\beta$ flux, and H$\beta$ equivalent width derived from VLT/X-shooter spectra. Panels (d)-(f) show the same but for H$\alpha$ and SALT spectra. For the continuum, flux was integrated over the spectrum masking the line regions, whereas only the lines themselves were integrated when obtaining the line flux. The solid line shows the spin frequency, whereas the dashed line shows beat frequency $\omega - \Omega$.}
\label{fig:linesFT}
\end{suppfigure*} 

\begin{suppfigure*}[h!]
\centering
\includegraphics[width=0.8\hsize]{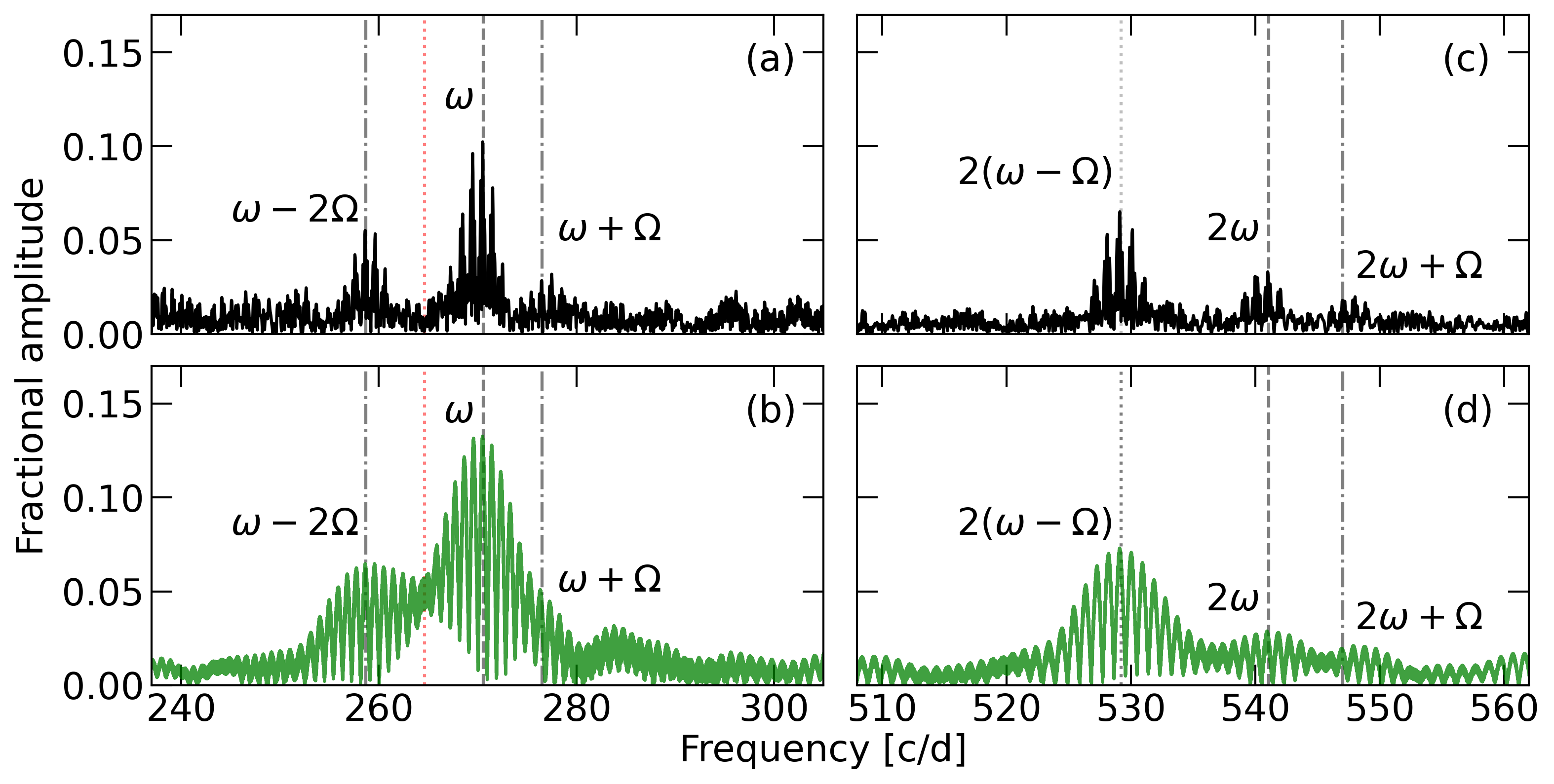}
\caption{{\bf Fourier transforms of HIPPO and ULTRACAM $g_s$.} The left panels show the Fourier transform of HIPPO data (a) and ULTRACAM $g_s$ (b) around the spin frequency $\omega$. Other peaks also visible are labelled. The beat frequency $\omega-\Omega$ (red dotted line) is not detected. The right panels show the same Fourier transforms around the first harmonic of the spin frequency, showing that the first harmonic of the beat frequency is detected and even dominant over the spin harmonic.}
\label{fig:FTbeat}
\end{suppfigure*}

\begin{suppfigure*}[h!]
\centering
\includegraphics[width=0.8\hsize]{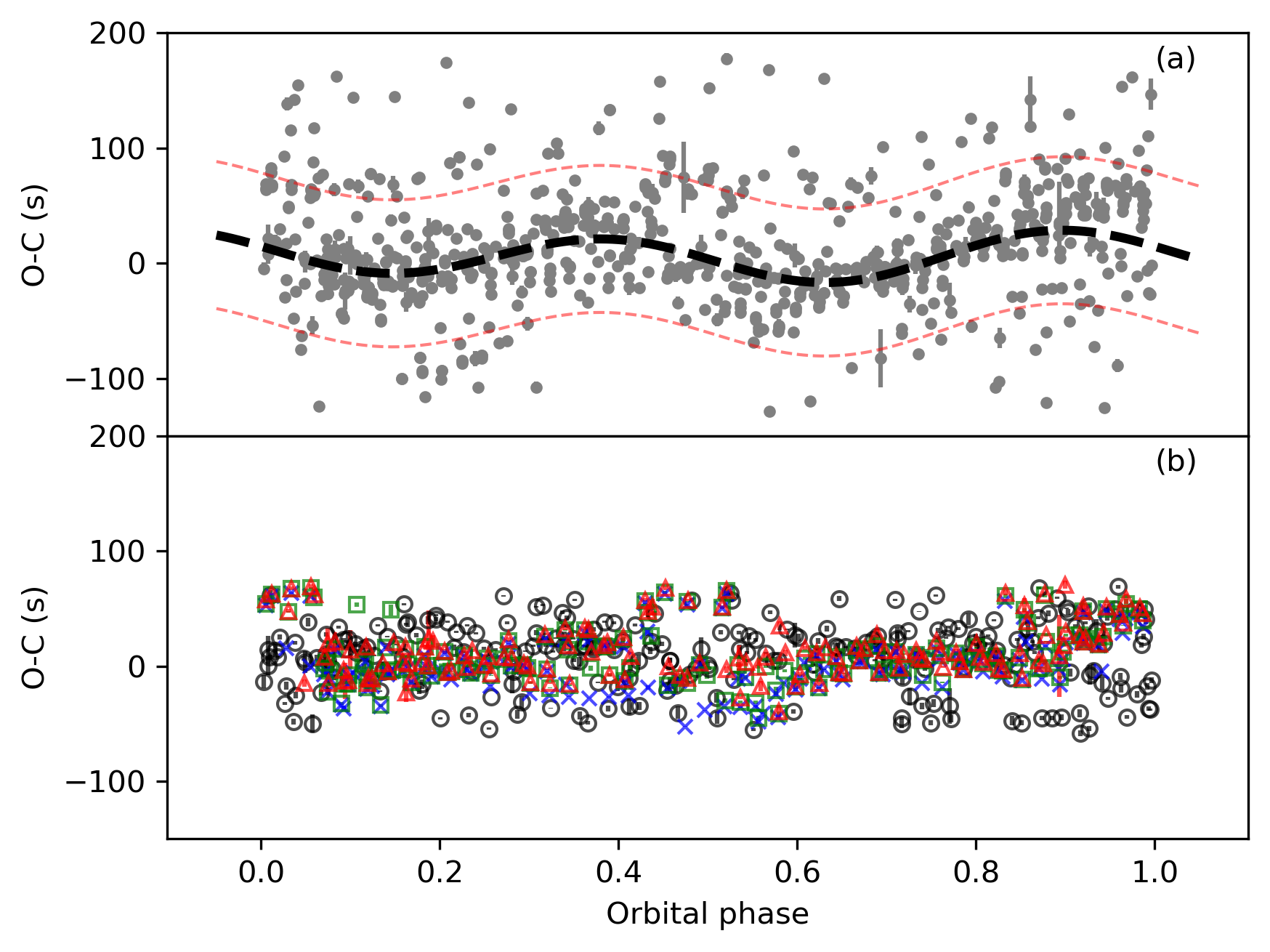}
\caption{{\bf Residuals from linear fit.} Panel (a) shows the residuals as a function of orbital phase after an initial linear fit to the pulse arrival times derived from ULTRACAM and HIPPO data. There is modulation with orbital phase, modelled by a Fourier series shown by the black dashed line. The red dashed lines show the same model plus or minus 0.25~cycles; points below or above the red lines were excluded from the final fit. Panel (b) shows the residuals after the fit that resulted in the ephemeris of Equation~\ref{spin_eph}. In panel (b) the symbols are grey circles for HIPPO data, red triangles for ULTRACAM $r_s$ and $i_s$ data, green squares for ULTRACAM $g_s$, and blue crosses for ULTRACAM $u_s$.}
\label{fig:omc}
\end{suppfigure*}

\begin{suppfigure*}[h!]
\centering
\includegraphics[width=0.8\hsize]{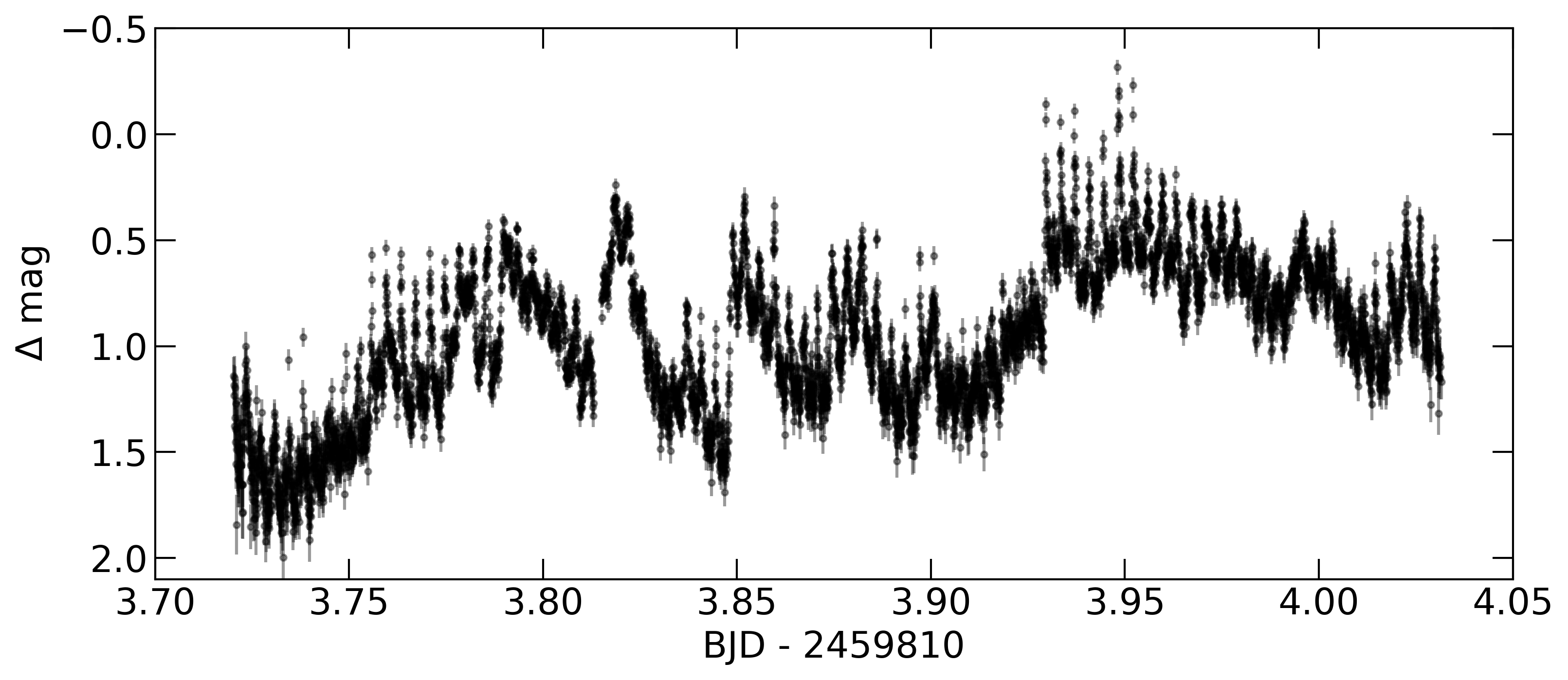}
\caption{{\bf Optical light curve showing possible flaring.} Light curve of \targ\ taken with the Mookodi instrument on the SAAO 1-m Lesedi telescope on 2022 August 21. There are strong changes in amplitude, e.g. between BJD 2459813.8 and 2459813.85, which are suggestive of flaring.
}
\label{fig:mookodi}
\end{suppfigure*}

\begin{suppfigure*}[h!]
\centering
\includegraphics[width=0.8\hsize]{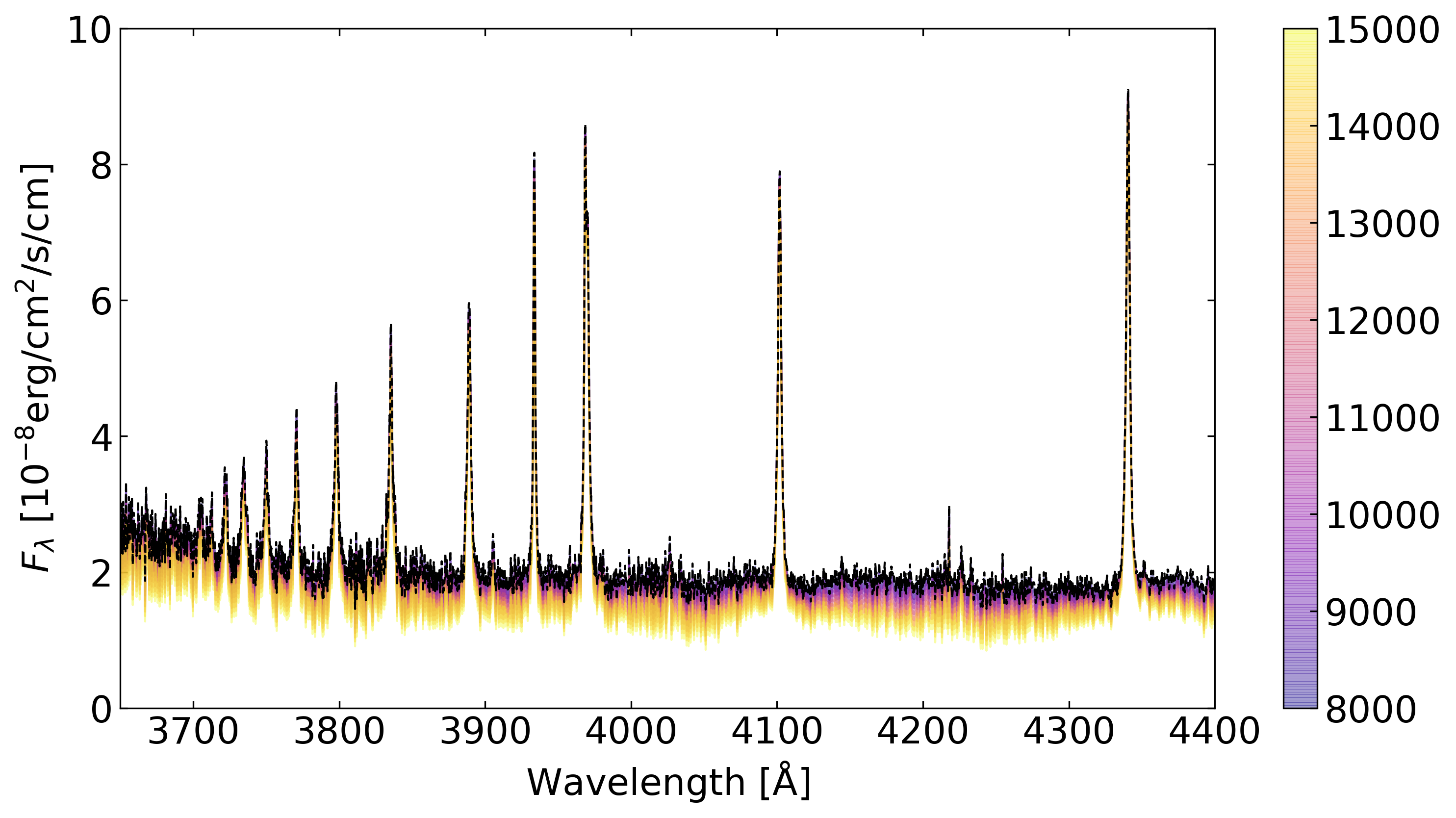}
\caption{{\bf Constraining the temperature of the white dwarf.} Panel (a) shows in grey a spectrum obtained by co-adding all UVB spectra taking around minimum flux. Model white dwarf spectra with $\log g = 8.50$ and different effective temperatures, indicated by the colour bar, are shown. The models were scaled taking \targ's distance into account and assuming a white dwarf radius from cooling models for $\log g = 8.50$. In panel (b), the model white dwarf spectra were subtracted from the observed spectrum. For the correct white dwarf model, this should simply remove the white dwarf contribution, and add no new features. From a certain temperature, the strength of the lines in the model is such that the subtraction visibly adds broad features to the spectrum. By numerically determining when that happens, we placed an upper limit on the white dwarf temperature of 13\,000~K.}
\label{wdteff}
\end{suppfigure*}

\begin{suppfigure*}[h!]
\centering
\includegraphics[width=\hsize]{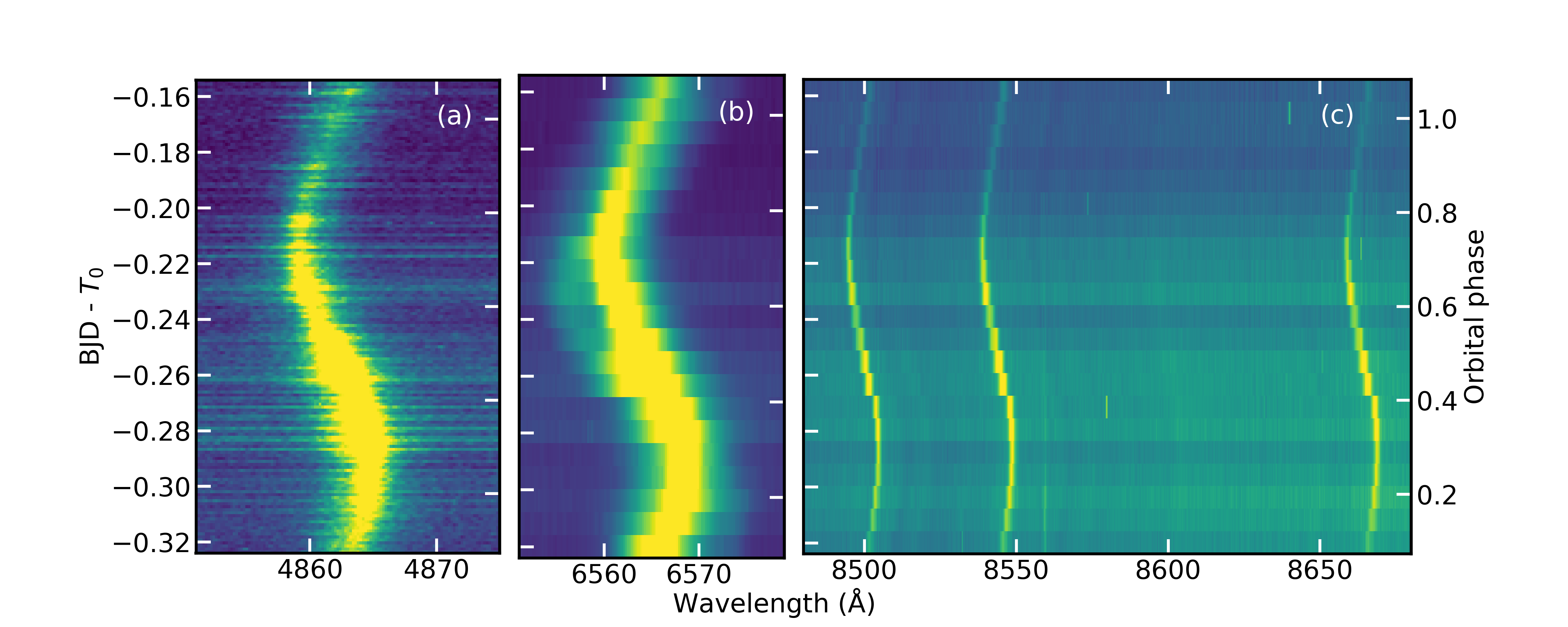}
\caption{{\bf Trailed spectra of \targ.} Panel (a) shows spectra taken with X-shooter's UVB arm, around H$\beta$. Panels (b) and (c) show data from the VIS arm, around H$\alpha$ and the CaII triplet, respectively. The left hand axis shows time in BJD, and the right hand axis shows orbital phases for all panels. For H$\beta$, the integration time was short enough to resolve the pulsations, which can be seen in the continuum.}
\label{trails}
\end{suppfigure*}

\begin{suppfigure*}[h!]
\centering
\includegraphics[width=0.7\hsize]{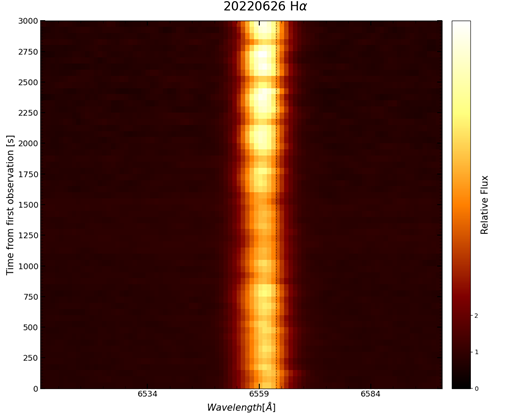}
\caption{{\bf Trailed H$\alpha$ spectra.} Trail of spectra obtained via frame-transfer spectroscopy with SALT. The pulsed behaviour of H$\alpha$ can be clearly seen in the flux changes, in particular towards the end of observations (top of the plot).}
\label{trail_alpha}
\end{suppfigure*}

\begin{suppfigure*}[h!]
\centering
\includegraphics[width=\hsize]{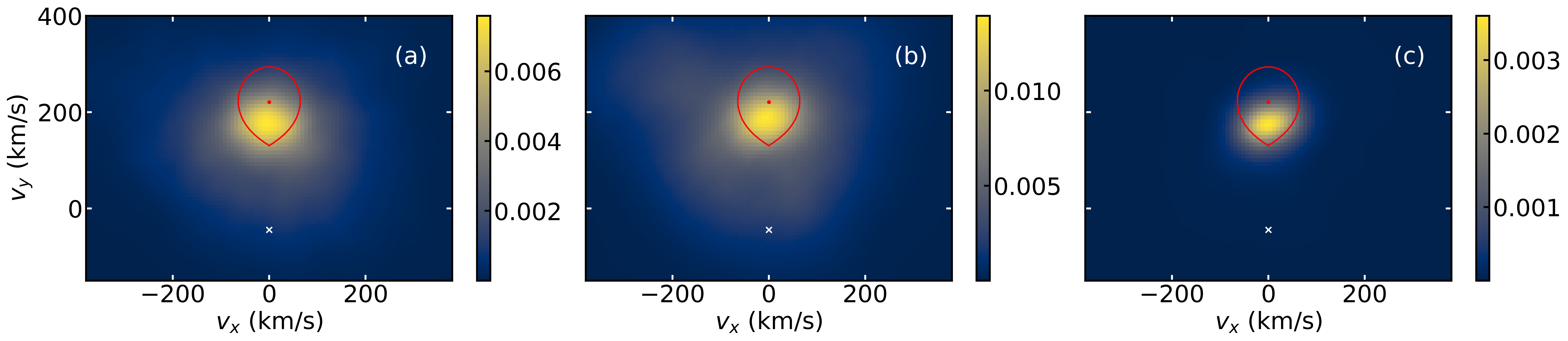}
\caption{{\bf Doppler tomography of \targ.} Doppler tomography of the H$\beta$ (panel a), H$\alpha$ (panel b) and Ca~II triplet (panel c) lines. The white cross marks the position of the white dwarf, the red line indicates the Roche lobe of the companion assuming $q=0.2$, and the red dot its centre of mass. The colour bars show emission intensity after normalising and subtracting the continuum.}
\label{doppler}
\end{suppfigure*}

\begin{suppfigure*}[h!]
\centering
\includegraphics[width=\hsize]{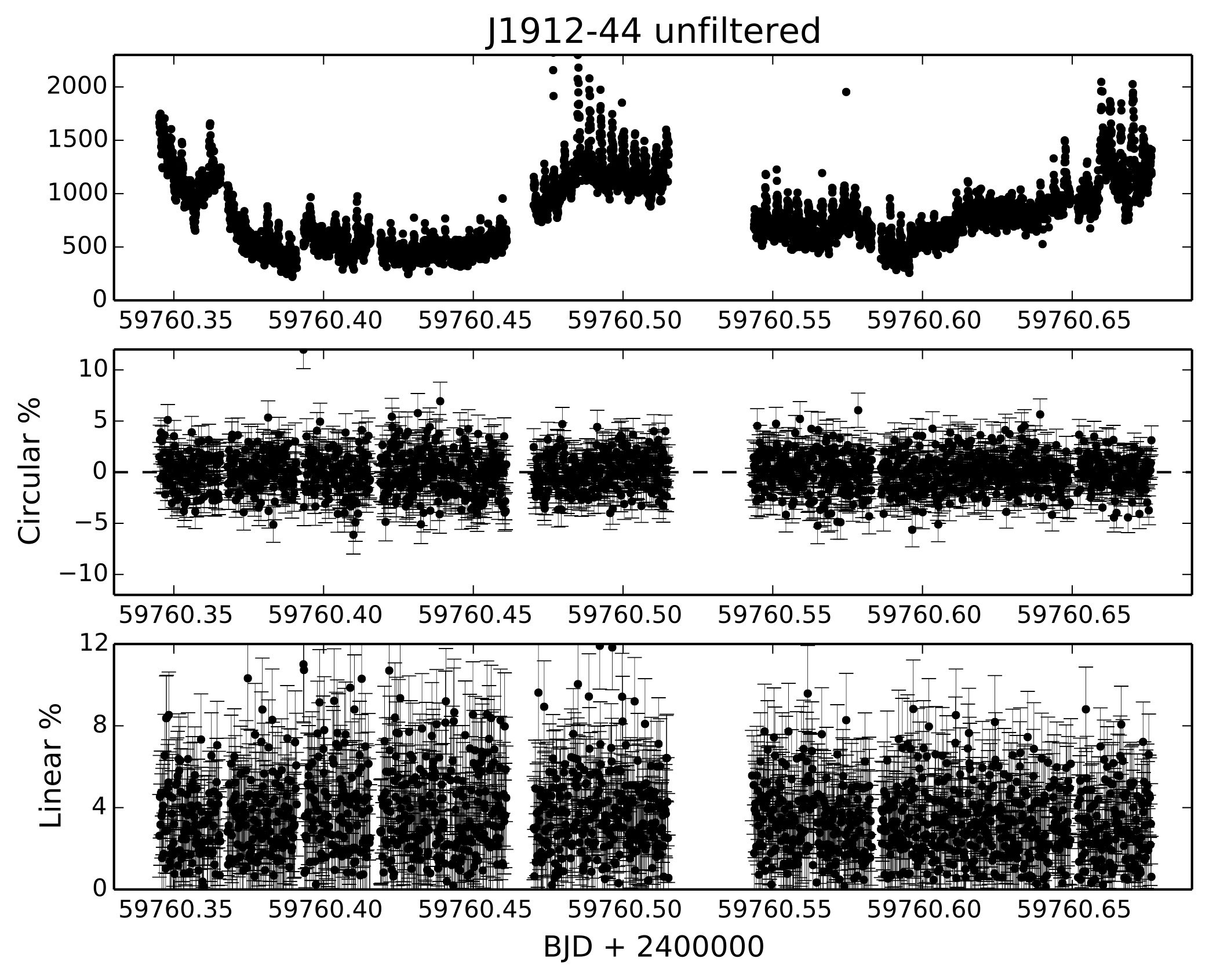}
\caption{{\bf Example HIPPO photo-polarimetry of \targ\ from 29 June 2022.} Top to bottom show the photometry, circular polarisation and linear polarisation, respectively, with one-sigma measurement uncertainties for each datapoint.}
\label{egHIPPO}
\end{suppfigure*}

\begin{suppfigure*}[h!]
\centering
\includegraphics[width=\hsize]{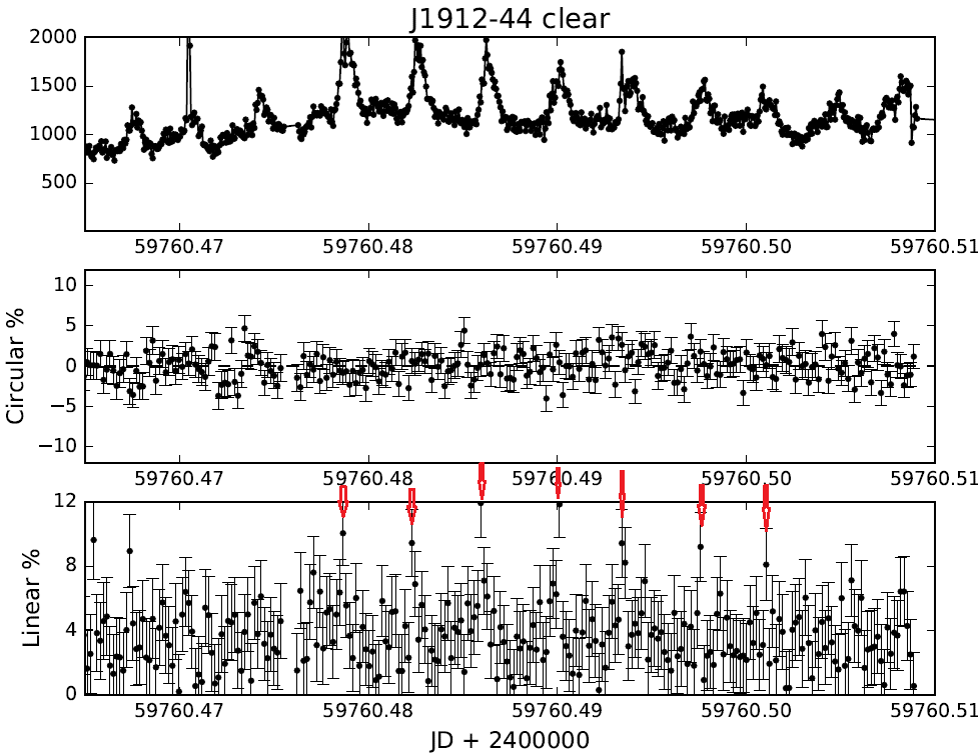}
\caption{{\bf Zoomed in region of Supplementary 
Figure~\ref{egHIPPO} highlighting the linear polarisation pulses.} Top to bottom show the photometry, circular polarisation and linear polarisation, respectively, with one-sigma measurement uncertainties for each datapoint..}
\label{egHIPPO2}
\end{suppfigure*}

\begin{suppfigure*}[h!]
\centering
\includegraphics[width=\hsize]{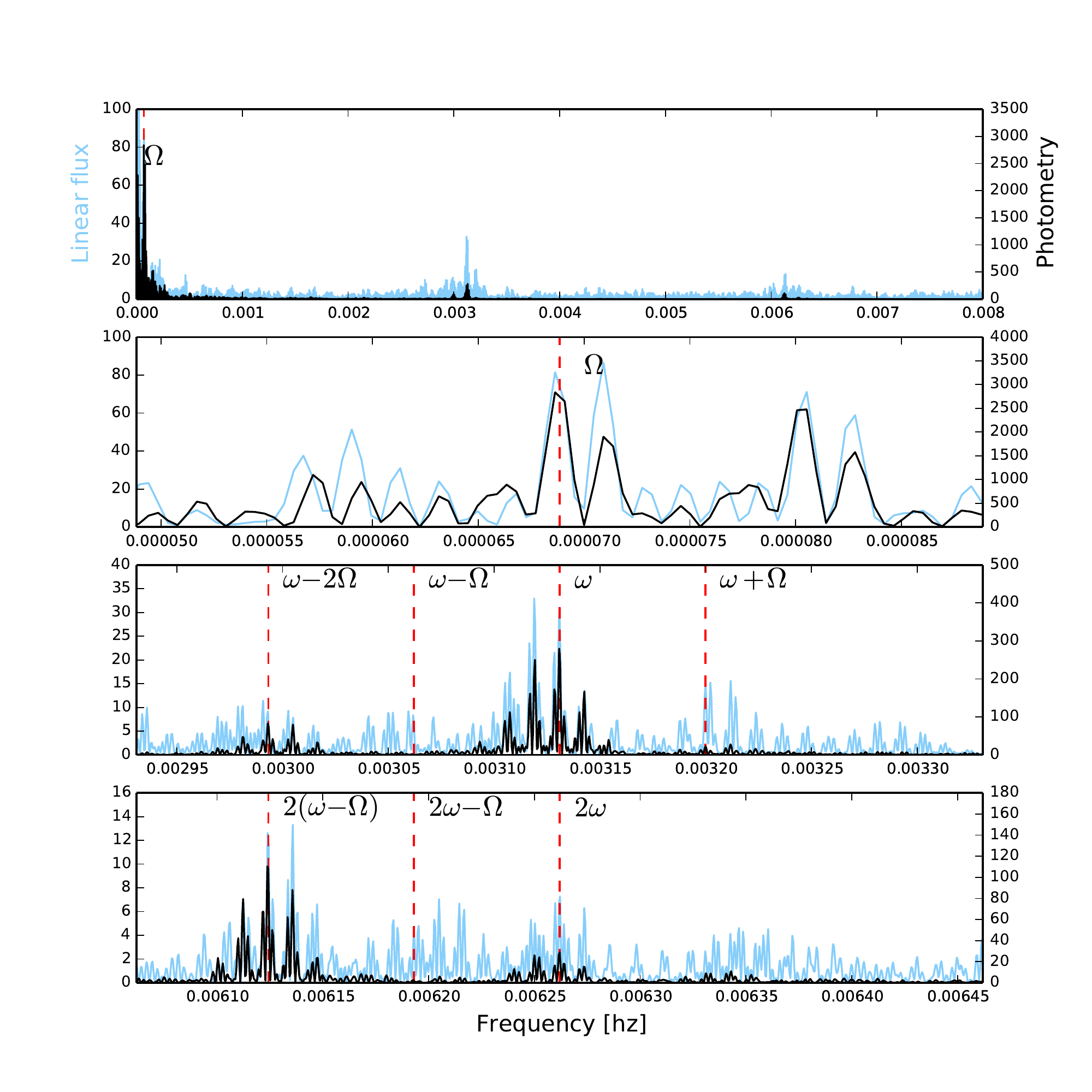}
\caption{{\bf Fourier transforms of HIPPO photometry and linear polarimetry.} Black is photometry, lightskyblue is linear polarised flux.}
\label{FTHIPPO}
\end{suppfigure*}

\begin{suppfigure*}[h!]
\centering
\includegraphics[width=0.9\hsize]{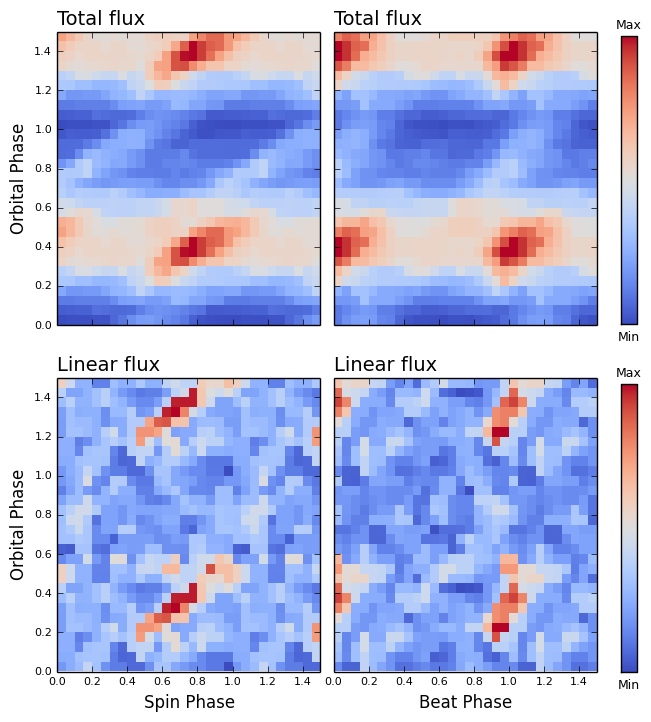}
\caption{{\bf Photopolarimetry of J1912-4410 assuming beat to be the dominant frequency.} Panels show the same as in Figure~\ref{fig:phot_pol}, but assuming beat rather than the spin as the dominant frequency, which results in considerable smearing in the spin panel. The right column is the same as the second column in Figure~\ref{fig:phot_pol}, because the same frequency is used for folding in both cases, but with a different interpretation.}
\label{fig:wrong_freqs}
\end{suppfigure*}

\begin{suppfigure*}[h!]
\centering
\includegraphics[width=0.9\hsize]{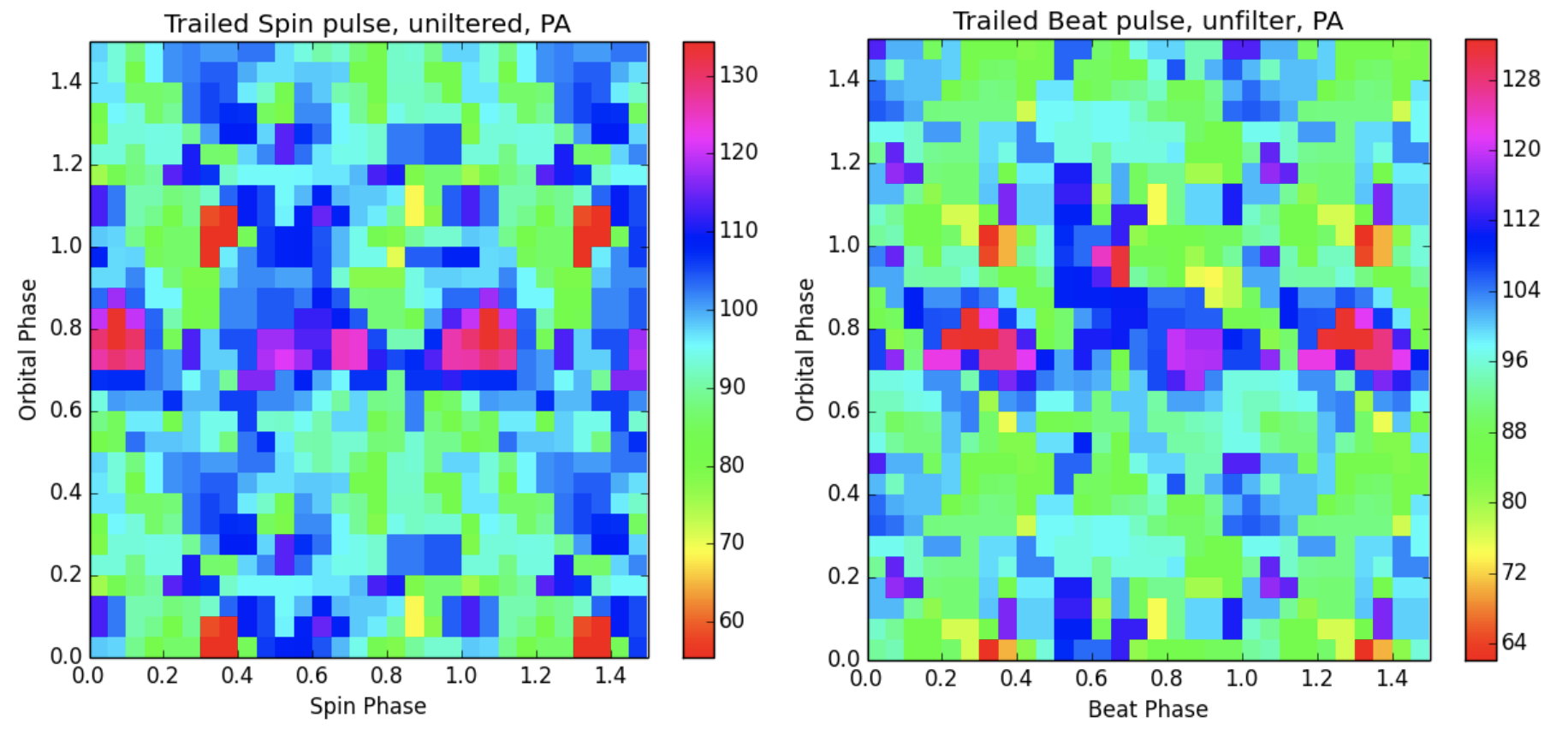}
\caption{{\bf Position angle of the linear polarisation as a function of orbital and spin/beat phases.} The left panel shows the position angle when the dominant frequency is assumed to be the spin, while the right shows the same when the beat is assumed to be dominant. In the left panel, there is a hint of a consistent angle for orbital phase $\approx 0.45$.}
\label{fig:polangle}
\end{suppfigure*}


\end{document}